\documentclass[aps,pre,
   twocolumn,
   preprint,
   tightenlines,
   reprint,
   superscriptaddress,
   eqsecnum,
   floatfix,
   showpacs,
   preprintnumbers,
   longbibliography,
   nofootinbib
   ]{revtex4-1}
\usepackage{color}
\usepackage{framed}
\usepackage{subfigure}
\usepackage{time}
\usepackage{graphicx}
\usepackage{enumerate}
\usepackage{latexsym}
\usepackage{bm}
\usepackage{upgreek}
\usepackage{amsthm}
\usepackage{amssymb}
\usepackage{mathrsfs}
\usepackage{amsmath}
\newcommand{\ef}[1]{\eqref{#1}}                
\usepackage{physics}                           
\newcommand{\PT}{\mathcal{PT}}
\newcommand{\PB}[2]{\ensuremath{ \lbrace \,#1,#2\, \rbrace }}  
\newcommand{\GammaF}[1]{\Gamma[\, #1 \,]}      
\newcommand{\tint}{\!\int\!}                   
\newcommand{\rmi}{{\rm i}}                     
\newcommand{\rme}{{\rm e}}                     
\newcommand{\tpsi}{\tilde{\psi}}               
\newcommand{\tphi}{\tilde{\phi}}               
\newcommand{\tb}{\tilde{b}}                    
\newcommand{\be}{\begin{equation}}
\newcommand{\bq}{\begin{equation}}
\newcommand{\ee}{\end{equation}}
\newcommand{\eq}{\end{equation}}
\newcommand{\bea}{\begin{eqnarray}}
\newcommand{\eea}{\end{eqnarray}}
\newcommand{\ba}{\begin{eqnarray}}
\newcommand{\ea}{\end{eqnarray}}
\newcommand{\vs}{vs.}                      
\newcommand{\ie}{\textit{i.e.}}            
\newcommand{\Schrodinger}{Schr{\"o}dinger} 
\usepackage{hyperref}
\hypersetup{
    unicode=false,          
    pdftoolbar=true,        
    pdfmenubar=true,        
    pdffitwindow=false,     
    pdfstartview={FitH},    
    pdftitle={Paper},       
    pdfauthor={John Dawson},     
    pdfsubject={Phys Rev article},   
    pdfcreator={John Dawson},   
    pdfproducer={John Dawson}, 
    pdfkeywords={keyword1} {key2} {key3}, 
    pdfnewwindow=true,      
    colorlinks=true,        
    linkcolor=red,          
    citecolor=blue,         
    filecolor=green,        
    urlcolor=cyan           
}
\begin{document}
%
%
\preprint{NLSE-Stability-v10.tex, \today, \now \ EDT}
\title{Stability of new exact solutions of the nonlinear \Schrodinger\ equation in a  P{\"o}schl-Teller  external potential } 
\author{John F. Dawson}
\email{john.dawson@unh.edu}
\affiliation{Department of Physics,
   University of New Hampshire,
   Durham, NH 03824, USA}   
\author{Fred Cooper} 
\email{cooper@santafe.edu}
\affiliation{The Santa Fe Institute, 1399 Hyde Park Road, Santa Fe, NM 87501, USA}
\affiliation{Theoretical Division and Center for Nonlinear Studies,
   Los Alamos National Laboratory,
   Los Alamos, NM 87545}
\author{Avinash Khare}
\email{khare@physics.unipune.ac.in}
\affiliation{Physics Department, Savitribai Phule Pune University, Pune 411007, India}
%
%
\author{Bogdan Mihaila}
\email{bmihaila@nsf.gov}
\affiliation{National Science Foundation, Arlington, VA 22230, USA}
\affiliation{
   Los Alamos National Laboratory,
   Los Alamos, NM 87545, USA}
\author{Edward Ar\'evalo}
\email{earevalo@fis.puc.cl} 
\affiliation{Pontifical Catholic University of Chile,
Departamento de F\'{\i}sica, Santiago, Regi\'on Metropolitana, Chile}

\author{Ruomeng Lan}
\email{rlan@math.tamu.edu}
\affiliation{Department of Mathematics, Texas A\&M University, 
             College Station, TX 77843, USA}

\author{Andrew Comech}
\email{comech@math.tamu.edu} 
\affiliation{Department of Mathematics, Texas A\&M University, 
             College Station, TX 77843, USA}
\affiliation{St. Petersburg State University, St. Petersburg 199178, Russia}  
\affiliation{Institute for Information Transmission Problems, Moscow 101447, Russia}

\author{Avadh Saxena} 
\email{avadh@lanl.gov}
\affiliation{Theoretical Division and Center for Nonlinear Studies,
   Los Alamos National Laboratory,
   Los Alamos, NM 87545}
\date{\today, \now \ EDT}
%
%
%
\begin{abstract}
We  discuss the stability properties of the solutions of the general nonlinear \Schrodinger\  equation (NLSE) in 1+1 dimensions in an external potential derivable from a parity-time ($\PT$) symmetric superpotential $W(x)$ that we considered earlier \cite{PhysRevE.92.042901}.  In particular we consider the nonlinear partial differential equation 
$  \{
   i \,
   \partial_t
   +  
   \partial_x^2
   - 
   V(x)
   + g 
   | \psi(x,t) |^{2\kappa}
   \} \, \psi(x,t) 
   = 
   0 \>, $
for arbitrary nonlinearity parameter $\kappa$, where $g= \pm1$ and $V$ is the well known P{\"o}schl-Teller potential which we allow to be repulsive as well as attractive.  Using energy landscape methods, linear stability analysis as well as a time dependent variational approximation, we derive consistent analytic results for the domains of instability of these new exact solutions as a function of the strength of the external potential and $\kappa$.  For the repulsive potential (and $g=+1$) we show that there is a translational instability which can be understood in terms of the energy landscape as a function of a stretching parameter and a translation parameter being a saddle near the exact solution. In this case,  numerical simulations show that if we start with the exact solution,  the initial wave function breaks into two pieces traveling in opposite directions. If we explore the slightly perturbed solution situations,  a 1\% change in initial conditions can change significantly the details of how the wave function breaks into two separate pieces. For the attractive potential (and $g=+1$), changing the initial conditions by 1 \% modifies the domain of stability only slightly.  
For the case of the attractive potential and negative $g$ perturbed solutions merely oscillate with the oscillation frequencies predicted by the variational approximation.  
\end{abstract}
\maketitle
%
%

%
%
\section{\label{s:Intro}Introduction}

The study of open systems with balanced loss and gain, typically defined by Parity-Time ($\PT$) symmetry, has elicited significant attention from physics, nonlinear science and 
mathematics communities during the past decade.  This is in part due to their emerging applications in many physical contexts and in part due to their intriguing mathematical structure. The initial investigation of such systems \cite{r:Bender:2007nr,0305-4470-39-32-E01,1751-8121-41-24-240301,1751-8121-45-44-440301} arose in the context of whether non-Hermitician quantum systems could lead to entirely real eigenvalues. Keeping in perspective the formal similarity of the \Schrodinger\ equation with Maxwell's equations in the paraxial approximation, many experimentalists realized that such $\PT$-invariant systems can indeed be fabricated using optical means \cite{Makris2011,0305-4470-38-9-L03,PhysRevLett.100.103904,PhysRevLett.101.080402,PhysRevLett.103.123601,PhysRevB.80.235102,PhysRevA.81.022102,r:Ruter:2010mz,PhysRevLett.103.093902,r:Regensburger:2012gf}. Motivated by this success, in the ensuing years, $\PT$-invariant phenomena was also observed in electronic circuits \cite{PhysRevA.84.040101,1751-8121-45-44-444029}, mechanical constructs \cite{r:Bender:2013ly}, whispering-gallery microcavities \cite{r:Peng:2014ul}, among many other physical systems.

In a parallel development, the concept of supersymmetry (SUSY) prevalent in high-energy physics was also experimentally studied in optics \cite{PhysRevLett.110.233902,r:Heinrich:2014qf}.  The underlying notion is that for a given potential we can obtain a SUSY partner potential such that both potentials possess identical spectrum (with possibly one eigenvalue different) \cite{0038-5670-28-8-R01,r:CooperKhareSukhatmePR}.  A simultaneous presence of $\PT$-symmetry and SUSY can lead to unexpected phenomena and is likely to be very  useful in achieving transparent and one-way reflectionless complex optical potentials \cite{0305-4470-33-1-101,doi:10.1142/S0217751X01004153,Bagchi2000285,Ahmed2001343,PhysRevA.89.032116}.  Previously \cite{PhysRevE.92.042901} we studied the interplay between nonlinearity, $\PT$-symmetry and supersymmetry as well as the rich consequences of this interplay. There we obtained exact solutions of the general nonlinear \Schrodinger\ equation (NLSE) in 1+1 dimensions in the presence of a $\PT$-symmetric complex potential \cite{0038-5670-28-8-R01,r:CooperKhareSukhatmeBOOK}.  In a recent paper \cite{Cooper:2017aa} we studied the stability properties of the solutions of NLSE in the real partner potential of the problem studied in \cite{PhysRevE.92.042901} which was a P{\"o}schl-Teller potential \cite{r:Poschl:1933ek,r:Landau:1989jt}.  

Here our objective is to discuss the stability properties of two related exact solutions which exist when we change the sign of the nonlinear coupling $g$ to being negative keeping the potential attractive, or keep the sign of the nonlinear term unchanged but consider a repulsive potential.   In the latter case we will find that the solutions are translationally unstable, whereas in the former case the solutions are stable to small perturbations

%
%
\subsection{\label{s:Model} Different solutions to the NLSE in an external P{\"o}schl-Teller potential}

By allowing the nonlinearity coupling $g=\pm 1$ and the sign of the potential $\lambda = \pm 1$ we have found different classes of exact soltuions when  the NLSE is in the presence of a P{\"o}schl-Teller potential centered at $x=0$.   \Schrodinger's equation for these cases is given by:
\begin{equation}\label{e:NLSE}
   \{\,
      \rmi \, \partial_t
      +
      \partial_{x}^2
      +
      g \, \abs{\psi(x,t)}^{2 \kappa}
      -
      V(x) \,
   \} \, \psi(x,t)
   =
   0 \>,
\end{equation}
where
\begin{equation}\label{e:NLSE-2}
  V(x)
  =
  - \lambda \, \tb^2 \, \sech^2(x)
  \qc
  \tb^2 = b^2 - 1/4 \>,
\end{equation}
with $\tb^2 > 0$ and $\kappa > 0$.  Since $g$ can be scaled out of the equation by letting $\psi(x,t) \mapsto  g^{-1/( 2\kappa)}\, \psi(x,t)$, we can restrict ourselves to $g = \pm 1$ in what follows.  The signs here are chosen such that the nonlinear term is attractive for $g = +1$ and  repulsive for $g = -1$ and the external P{\"o}schl-Teller potential is attractive for $\lambda = +1$ and repulsive for $\lambda = -1$.  This potential is a special case of potentials obtainable from the complex $\PT$-symmetric SUSY superpotential 
\begin{equation}\label{e:NLSE-3}
   W(x)
   =
   \qty( m - 1/2 ) \,  \tanh{x} 
   - 
   i b \, \sech{x} \>,
\end{equation}
with $m=1$, which gives rise to $\PT$-symmetric partner potentials $V_{\pm} = W^2  \pm W'$. Our real $V(x)$ corresponds to $V_{+}$.  There are several cases of Eq.~\ef{e:NLSE} which have exact solutions.  These are
\begin{enumerate}[(I)]
\item Attractive nonlinear term and attractive potential: $g = +1$, $\lambda = +1$.  In this case, the exact solution is given by
\begin{subequations}\label{e:NLSE-4}
\begin{align}
   \psi_0(x,t)
   &=
   A_0(\tb,\gamma) \, \sech^{\gamma}(x) \, \rme^{\rmi \gamma^2 t}
   \label{e:NLSE-4a} \\
   A_0^{2/\gamma}(\tb,\gamma)
   &=
   \gamma ( \gamma + 1 ) - \tb^2 \>,
   \label{e:NLSE-4b}
\end{align}
\end{subequations}
where $\gamma = 1/\kappa$.  In this case,
\begin{equation}\label{e:NLSE-4.1}
   \tb^2_{\gamma} 
   \equiv
   \gamma ( \gamma + 1 ) 
   \ge 
   \tb^2 \ge 0 \>.
\end{equation}
We studied this case in a previous paper \cite{Cooper:2017aa}, where we found that all solitary waves for $\kappa < 2$ and $0 < \tb^2 < \tb^2_{\gamma}$ are stable, as for the case of solitary waves in the NLSE ($\tb^2=0$).  However, we also found a new region above $\kappa=2$ where these solutions are stable.
\item Attractive nonlinear term and repulsive potential: $g = +1$, $\lambda = -1$. 
For this case, the exact solution is given by
\begin{subequations}\label{e:NLSE-5}
\begin{align}
   \psi_0(x,t)
   &=
   A_0(\tb,\gamma) \, \sech^{\gamma}(x) \, \rme^{\rmi \gamma^2 t}
   \label{e:NLSE-5a} \\
   A_0^{2/\gamma}(\tb,\gamma)
   &=
   \gamma ( \gamma + 1 ) + \tb^2 \>.
   \label{e:NLSE-5b}
\end{align}
\end{subequations}
In this case, we only require $\tb^2 \ge 0$.
This solution goes over to a particular moving solitary wave  solution of the NLSE when $\tb \rightarrow 0$. Since the solutions of the NLSE are stable to deformations of the width for all $\kappa < 2$ we expect (and we will find) that in that regime there will be a critical value of $\tb$ above which the solution will be \emph{unstable} to width deformations.  We expect and we find that again the solutions are always unstable for $\kappa > 2$.  What we will also find is that for all values of $\kappa$ these solutions are \emph{unstable} to a slight translation, even if induced by numerical noise. 

\item Repulsive nonlinear term and attractive potential: $g = -1$, $\lambda = +1$.
For this case, the exact solution is given by
\begin{subequations}\label{e:NLSE-6}
\begin{align}
   \psi_0(x,t)
   &=
   A_0(\tb,\gamma) \, \sech^{\gamma}(x) \, \rme^{\rmi \gamma^2 t}
   \label{e:NLSE-6a} \\
   A_0^{2/\gamma}(\tb,\gamma)
   &=
   \tb^2 - \gamma ( \gamma + 1 )  \>.
   \label{e:NLSE-6b}
\end{align}
\end{subequations}
In this case, we require $\tb^2 \ge \tb^2_{\gamma}$.
For this choice of $g$ there are no solitary wave solutions in the absence of the potential.  We will find that these solutions are linearly stable. %
\end{enumerate}
In all these cases, we find that the quantity
\begin{equation}\label{e:NLSE-7}
   g \, A_0^{2/\gamma}(\tb,\gamma)
   =
   \gamma ( \gamma + 1 ) - \lambda \, \tb^2 \>,
\end{equation}
is \emph{independent} of $g$  and depends only on the sign of $\lambda$.  The normalization, or ``mass'' of these exact wave functions is given by
\begin{equation}\label{e:NLSE-8}
   M_0(\tb,\gamma)
   =
   \tint \dd{x} \abs{\psi_0(x,t)}^2
   =
   A_0^{2}(\tb,\gamma) \, c_1[\gamma] \>,
\end{equation}
where
\begin{equation}\label{e:NLSE-9}
   c_1[\gamma]
   =
   \tint \dd{z} \sech^{2\gamma}(z)
   =
   \frac{\sqrt{\pi} \, \GammaF{\gamma}}{\GammaF{\gamma + 1/2}} \>.
\end{equation}
%

This paper is structured as follows: in Section~\ref{s:TDvar} we discuss Hamilton's principle of least action and the time-dependent variational approximation.  In Section~\ref{s:Derrick}, we use Derrick's theorem to study the stability of these solutions to width instabilities. In Section~\ref{s:translation} we discuss the energy landscape when we include translations of the origin of the solution. In Section~\ref{s:Linear} we perform a linear stability analysis.  In Section~\ref{s:fourtrial}, we introduce a four-parameter trial wave function to study the dynamics of the model, and in Section~\ref{s:numerics} we provide results of the direct numerical solutions of the nonlinear \Schrodinger\ equation in the P{\"o}schl-Teller  external potential.  Our main conclusions are summarized in Section~\ref{s:conclude}.

%
%
\section{\label{s:TDvar}Time-dependent variational principle}

The time-\emph{dependent} version of the variational approximation can
be traced to an obscure appendix in the 1930 Russian edition of the
``Principles of Wave Mechanics,'' by Dirac.\footnote{P.~A.~M.~Dirac,
Appendix to the Russian edition of \emph{The Principles of Wave
Mechanics}, as cited by Ia.~I.~Frenkel, \emph{Wave Mechanics, Advanced
General Theory} (Clarendon Press, Oxford, 1934), pp. 253, 436.  
Pattanayak and Schieve \cite{PhysRevE.50.3601} point out that the
reference often quoted, P.~A.~M.~Dirac, Proc.\ Cambridge Philos.\
Soc.\ \textbf{26}, 376 (1930), does not contain this
equation.}  In this version of the variational approximation, the wave
function is taken to be a function of a number of time-dependent
parameters.  Variation of the action, as defined by Dirac, leads to a
classical set of Hamiltonian equations of motion for the parameters.
These classical equations are then solved as a function of time to
provide an approximation to the evolution of the wave function.

The action which leads to Eq.~\eqref{e:NLSE-1} is given by 
\begin{equation}\label{e:HP-1}
   \Gamma[\psi,\psi^{\ast}] 
   = 
   \tint \dd{t} L[\psi,\psi^{\ast}] 
\end{equation}
where
\begin{subequations}\label{HP-2}
\begin{align}
   L[\psi,\psi^{\ast}]
   &=
   \frac{\rmi}{2} \tint \dd{x}
   \qty[\, 
      \psi^{\ast} (\partial_t \psi) 
      - 
      (\partial_t \psi^{\ast} ) \psi \,]
   -
   H[\psi,\psi^{\ast}] \>,
   \label{HP-2a} \\
   H[\psi,\psi^{\ast}]
   &=
   \tint \dd{x} 
   \bigl [ \,
       \abs{\partial_x \psi}^2
       -
       \frac{g \, \abs{\psi}^{2\kappa+2}}{\kappa + 1}
       +
       V(x) \, \abs{\psi}^2 \,
    \bigr ] \>.
    \label{HP-2b}
\end{align}
\end{subequations}
The NLSE and its complex conjugate follow from minimizing the action via,
\begin{equation}\label{HP-3}
   \frac{\delta \Gamma}{\delta \psi^\ast} 
   = 
   \frac{\delta \Gamma}{\delta \psi} 
   = 0 \>.
\end{equation}

%
%
\subsection{\label{ss:symplectic}Symplectic formulation}

In this section it will be useful to introduce a symplectic formulation of Lagrange's equations for the variational parameters.
We consider a variational wave function of the form:
\begin{equation*}
   \tilde{\psi}[\,x,Q(t)\,] 
   \qc
   Q(t) = \{\, Q^1(t),Q^2(t),\dotsc,Q^{2n}(t) \,\} \>.
\end{equation*}
Introducing the notation $\partial_i \equiv \partial / \partial Q^i$, the Lagrangian \ef{HP-2a} is given by
\begin{equation}\label{e:VT-2}
   L[\,Q,\dot{Q}\,]
   =
   \pi_i(Q) \, \dot{Q}^i - H[\,Q\,] \>,
\end{equation}
where
\begin{equation}\label{e:VT-3}
   \pi_i(Q)
   =
   \frac{\rmi}{2} \tint \dd{x}
   \{ \, 
      \tpsi^{\ast}\,[\, \partial_i \tpsi \, ]
      - 
      [\, \partial_i \tpsi^{\ast} \,] \, \tpsi
   \,\} \>,
\end{equation}
and $H(Q)$ is given by
\begin{equation}\label{e:VT-4}
   H(Q)
   =
   \tint \dd{x} 
   \bigl [ \,
       |\partial_x \tpsi |^2
       -
       \frac{g \, |\tpsi|^{2\kappa+2}}{\kappa + 1}
       +
       V(x) \, |\tpsi|^2 \,
    \bigr ] \>.   
\end{equation}
The Euler-Lagrange equations now become
\begin{equation}\label{e:VT-6}
   \dv{t} \Bigl ( \pdv{L}{\dot{Q}^i} \Bigr )
   -
   \pdv{L}{Q^i}
   =
   0 \>.
\end{equation}
From \ef{e:VT-2} this gives
\begin{equation}\label{e:VT-7}
   f_{ij}(Q) \, \dot{Q}^j
   =
   v_i(Q) \>,
\end{equation}
where we have set $v_i(Q) \equiv \partial_i H(Q)$, and where
\begin{equation}\label{e:VT-8}
   f_{ij}(Q)
   =
   \partial_i \pi_j(Q) - \partial_j \pi_i(Q)
\end{equation}
is an antisymmetric $2n \times 2n$ symplectic matrix.  
If $\det{f(Q)} \ne 0$, we can define an inverse as the contra-variant matrix with upper indices,
\begin{equation}\label{e:VT-9}
   f^{ij}(Q) f_{jk}(Q) = \delta^i_k \>,
\end{equation}
in which case the equations of motion \ef{e:VT-7} can be put in the form:
\begin{equation}\label{e:VT-10}
   \dot{Q}^i
   =
   f^{ij}(Q) \, v_j(Q) \>.
\end{equation}
Conservation of energy is expressed as
\begin{equation}\label{e:VT-11}
   \dv{H(Q)}{t}
   =
   \dot{Q}^i \, v_i(Q)
   =
   f^{ij}(Q) \, v_j(Q) \, v_i(Q) 
   =
   0 \>,
\end{equation}
since $f^{ij}(Q)$ is an antisymmetric tensor.  Poisson brackets are defined using this antisymmetric tensor.  If $A(Q)$ and $B(Q)$ are functions of $Q$, Poisson brackets are defined by
\begin{equation}\label{e:VT-12}
   \PB{A(Q)}{B(Q)}
   =
   ( \partial_i A(Q) ) \, f^{ij}(Q) \, ( \partial_j B(Q) ) \>.
\end{equation}
In particular,
\begin{equation}\label{e:VT-13}
   \PB{Q^i}{Q^j}
   =
   f^{ij}(Q) \>.
\end{equation}
This definition satisfies Jacobi's identity.  That is, what we have shown here is that the $2n$ quantities $Q^i$ are classical symplectic variables.   

%
%
\section{\label{s:Derrick}Derrick's theorem}

%
%
\begin{figure*}[t]
  \centering
  \subfigure[\ Width stable regions for cases I and III.  The upper curve is 
  $\tb^2_{\gamma}$ from Eq.~\ef{e:NLSE-4.1}.  The lower curve is 
  $\tb^2_{\text{crit}}$ with $\lambda = +1$ from Eq.~\ef{e:DT-12}.]
  { \label{f:Fig1a} \includegraphics[width=0.95\columnwidth]{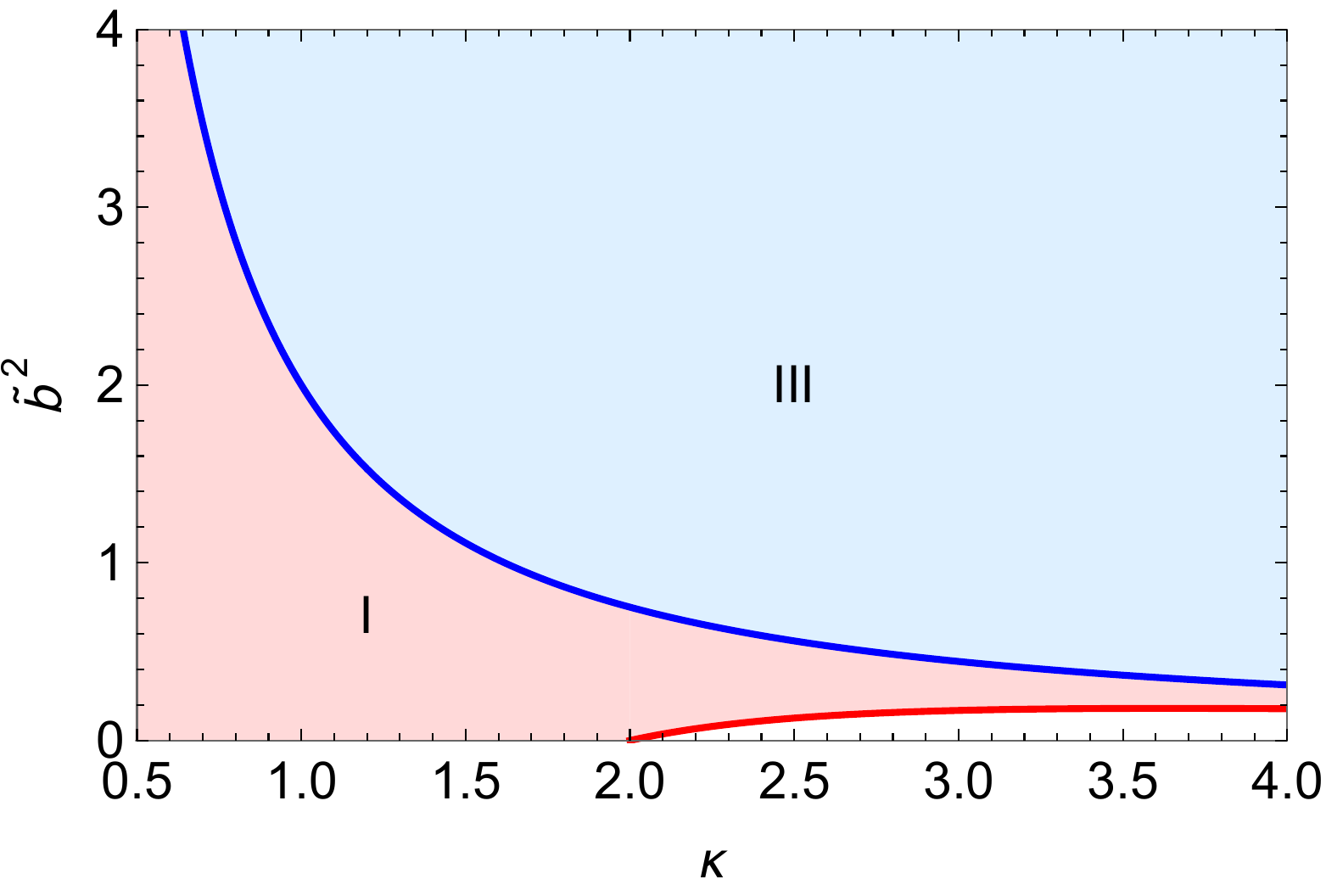} }
  \subfigure[\ Width stable region for case II.  The curve is 
  $\tb^2_{\text{crit}}$ with $\lambda = -1$ from Eq.~\ef{e:DT-12}.]
  { \label{f:Fig1b} \includegraphics[width=0.95\columnwidth]{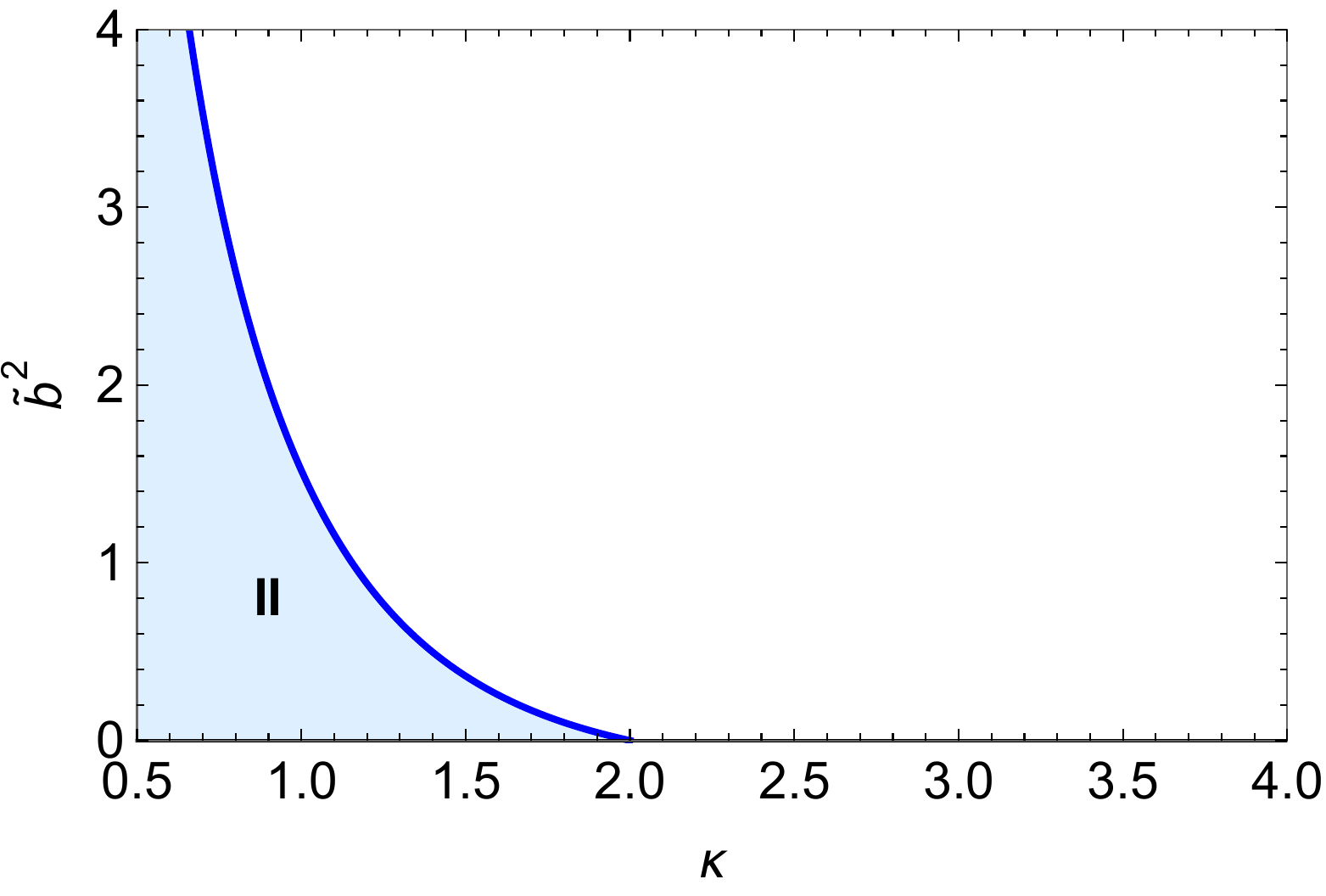} }
  \caption{\label{f:Fig1} Width stable regions for cases I, II, and III,
  according to Derrick's theorem.}
\end{figure*}
%
%

Derrick's theorem \cite{doi:10.1063/1.1704233} states that for a Hamiltonian dynamical system, an exact solution of the equation of motion  is unstable if under a scale transformation, $x \mapsto \beta x$ with fixed normalization, the energy of the system is lowered.   The stretched  wave function for Derrick's theorem is given by
\begin{equation}\label{e:DT-1}
   \psi_{\beta}(x,t)
   =
   A(\tb,\beta,\gamma) \, \sech^{\gamma}( \beta x) \>,
\end{equation}
with the normalization fixed by the requirement,
\begin{align}\label{e:DT-2}
   M[\tb,\beta,\gamma]
   &=
   \tint \dd{x} \abs{\psi_{\beta}(x,t)}^2
   =
   A^2(\tb,\beta,\gamma) \, c_1[\gamma] / \beta
   \\
   &= 
   M_0[\tb,\gamma]
   =
   A_0^2(\tb,\gamma) \, c_1[\gamma] \>.
   \notag
\end{align}
So $A^2(\tb,\beta,\gamma) = \beta A_0^2(\tb,\gamma)$.  Evaluation of the Hamiltonian \ef{HP-2b} with Derrick's wave function gives:
\begin{equation}\label{e:DT-3}
   H(\beta,\gamma)
   =
   H_1(\beta,\gamma) + H_2(\beta,\gamma) + H_3(\beta,\gamma) \>,
\end{equation}
where
\begin{subequations}\label{e:DT-4}
\begin{align}
   H_1(\beta,\gamma)
   &=
   \tint \dd{x} \abs{\partial_x \psi_{\beta}}^2
   \label{e:DT-4a} \\
   &=
   A_0^2 \, \beta^2 \gamma^2
   \tint \dd{z} \sech^{2\gamma+2}(z) \sinh^2(z)
   \notag \\
   &=
   \frac{A_0^2 \, \beta^2 \, \gamma}{2} \, c_1[\gamma+1] \>,
   \notag \\
   H_2(\beta,\gamma)
   &=
   - 
   \frac{g}{\kappa + 1}
   \tint \dd{x} \abs{\psi_{\beta}}^{2\kappa+2}
   \label{e:DT-4b} \\
   &=
   -
   g \, A_0^{2/\gamma} \, A_0^2 \,
   \frac{\gamma \, \beta^{1/\gamma}}{\gamma + 1}
   \tint \dd{z} \sech^{2 \gamma + 2 }(z) \>,
   \notag \\
   &=
   - 
   A_0^2 \,
   \frac{\gamma \, \beta^{1/\gamma}}{\gamma + 1} \,
   [\, \gamma ( \gamma + 1 ) - \lambda \, \tb^2 \,] \,
   c_1[\gamma+1] \>,
   \notag \\
   H_3(\beta,\gamma)
   &=
   \tint \dd{x} V(x) \, \abs{\psi_{\beta}}^2 
   \label{e:DT-4c} \\
   &=
   - \lambda \, \tb^2 \, A_0^2 \, \beta
   \tint \dd{x} \sech^{2}(x) \sech^{2\gamma}(\beta x)
   \notag \\
   &=
   - \lambda \, \tb^2 \, A_0^2 \, g_1[\beta,\gamma] \>,
   \notag
\end{align}
\end{subequations}
where
\begin{equation}\label{e:DT-5}
   g_1[\beta,\gamma]
   =
   \tint \dd{z} \sech^{2\gamma}(z) \, \sech^{2}(z/\beta) \>.
\end{equation}
So then $H(\beta,\gamma)/M_0[\tb,\gamma]$ is given by
\begin{align}
   &h(\tb,\beta,\gamma)
   \equiv
   \frac{H(\beta,\gamma)}{A_0^2(\tb,\gamma) \, c_1[\gamma+1]}
   \label{e:DT-6} \\
   &=
   \frac{1}{2} \, \beta^2 \, \gamma 
   -
   \frac{\gamma \, \beta^{1/\gamma}}{\gamma + 1} \,
   [\, \gamma ( \gamma + 1 ) - \lambda \, \tb^2 \,] \,
   - 
   \lambda \, \tb^2 \, \frac{g_1[\beta,\gamma]}{c_1[\gamma+1]} \>,
   \notag
\end{align}
and is independent of $g$.  $h(\tb,\beta,\gamma)$ is stationary when $\beta=1$.
We have
\begin{align}\label{e:DT-7}
   &\pdv{h(\tb,\beta,\gamma)}{\beta}
   =
   \gamma \,
   [\, \beta - \beta^{1/\gamma-1} \,]
   \\
   & \qquad
   +
   \lambda \, \tb^2 \,
   \qty[\,
      \frac{\beta^{1/\gamma-1}}{\gamma + 1}
      -
      \frac{1}{c_1[\gamma+1]} \, \pdv{g_1[\beta,\gamma]}{\beta} \,
   ] . 
   \notag
\end{align}
From Eq.~\ef{e:IT-4} in Appendix~\ref{s:integrals}, we find
\begin{equation}\label{e:DT-8}
   \pdv{g_1[\beta,\gamma]}{\beta} \Big |_{\beta=1}
   =
   \frac{c_1[\gamma+1]}{\gamma+1} \>,
\end{equation}
so that at $\beta = 1$, 
\begin{equation}\label{e:DT-9}
   \pdv{h(\tb,\beta,\gamma)}{\beta} \Big |_{\beta=1}
   =
   0 \>,
\end{equation}
for all values of $\gamma$ and $\tb^2$.  The sign of the second derivative of $h(\tb,\beta,\gamma)$ with respect to $\beta$ at $\beta=1$  determines whether the solution is unstable to small changes in the width. If $\partial^2 h(\tb,\beta,\gamma)/ \partial \beta^2$ evaluated at $\beta=1$ is negative the solution is unstable. We find
\begin{align}\label{e:DT-10}
   &\pdv[2]{h(\tb,\beta,\gamma)}{\beta}
   =
   \gamma + (\gamma-1) \, \beta^{1/\gamma - 2}
   \\
   & \quad
   -
   \lambda \, \tb^2 \,
   \qty[\,
      \frac{\gamma-1}{\gamma(\gamma+1)} \, \beta^{1/\gamma-2}
      +
      \frac{1}{c_1[\gamma+1]} \, \pdv[2]{g_1[\beta,\gamma]}{\beta} \,
   ] \>.
   \notag   
\end{align}
From Appendix~\ref{s:integrals}, we have
\begin{equation*}
   \pdv[2]{g_1[\beta,\gamma]}{\beta} \Big |_{\beta=1}
   =
   4 \, c_2[\gamma+1] 
   - 
   6 \, c_2[\gamma+2] 
   - 
   \frac{2 \, c_1[\gamma+1]}{\gamma+1} \>,
\end{equation*}
where
\begin{align}\label{e:DT-10.1}
   c_2[\gamma]
   &=
   \tint \dd{z} z^2 \, \sech^{2\gamma}(z)
   \\
   &=
   2^{2\gamma - 1}\,
   {}_4F_3[\gamma,\gamma,\gamma,2\gamma;1+\gamma,1+\gamma,1+\gamma;-1] / \gamma^3 \>.
   \notag    
\end{align}
Inserting this into \ef{e:DT-10} and evaluating it at $\beta=1$ gives:
\begin{align}\label{e:DT-11}
   &\pdv[2]{h(\tb,\beta,\gamma)}{\beta} \Big |_{\beta=1}
   =
   2 \gamma - 1 
   \\
   & \qquad
   +    
   \lambda \, \tb^2 \,
   \qty[\,
      \frac{1}{\gamma}
      -
      \frac{2\gamma+1}{\gamma} \,
      \frac{2 \, c_2[\gamma+1] - 3 \, c_2[\gamma+2]}{c_1[\gamma]} \,
       ] \>,
   \notag
\end{align}
which is independent of $g$.  

The critical value of $\tb^2$, when \ef{e:DT-11} vanishes, is 
\begin{equation}\label{e:DT-12}
   \tb^2_{\text{crit}}
   =
   - \lambda \,
   \frac{\gamma(2\gamma-1)}
   {1 
    - 
    (2\gamma+1) \,
    \frac{\displaystyle 2\, c_2[\gamma+1] - 3\, c_2[\gamma+2]}
         {\displaystyle c_1[\gamma]}  } \>,
\end{equation}
which is independent of $g$.  One can easily check that
\begin{equation}\label{e:DT-13}
    1 
    - 
    (2\gamma+1) \,
    \frac{2\, c_2[\gamma+1] - 3\, c_2[\gamma+2]}{c_1[\gamma]}
    > 0 \>,
\end{equation}
for all $\gamma$. In Appendix~\ref{s:integrals}, we give an alternative form for Eq.~\ef{e:DT-12}, which is in agreement with Ref.~[\onlinecite{Cooper:2017aa}].  In Fig.~\ref{f:Fig1}, we plot $\tb^2_{\text{crit}}$ for the three cases.  Referring to the figure, according to Derrick's theorem,

\begin{enumerate}[(I)]
\item For case I with $g=+1$, $\lambda = +1$, and $0 \le \tb^2 \le \tb^2_{\gamma}$, we see from Eq.~\ef{e:DT-11} that $\partial^2 h(\tb,\beta,\gamma)/\partial \beta^2 \ge 0$ for all $\kappa < 2$, so Derrick's theorem predicts that solutions are width stable for $\kappa < 2$.  For $\kappa > 2$, there is another region for $\tb^2_{\text{crit}} < \tb^2 < \tb^2_{\gamma}$ where width stable solutions are also possible.  
\item For case II with $g=+1$, $\lambda = -1$, and $0 \le \tb^2 \le \tb^2_{\text{crit}}$, width stable solutions are possible for $\kappa < 2$, but we will find that this region is unstable to translation perturbations.  

\item For case III with $g = -1$, $\lambda = +1$, and $\tb^2_{\gamma}< \tb^2$ width stable solutions are possible for all $\kappa$.
\end{enumerate}

Derrick's theorem is a version of the time-\emph{independent} variational method and only provides information about the stability of the system under a change in $\beta$, the width of the wave function.  Thus Derrick's theorem only gives a sufficient condition for an instability to occur.  To see if there are translational instabilities as well as width instabilities we will consider what happens to the energy when we make a small translation of the position of the solution.

%
%
\section{\label{s:translation} Translational Stability Landscape}

%
%
\begin{figure*}[t]
  \centering
  \subfigure[\ Case I, attractive potential.]
  { \label{f:Fig3Da} \includegraphics[width=0.95\columnwidth]{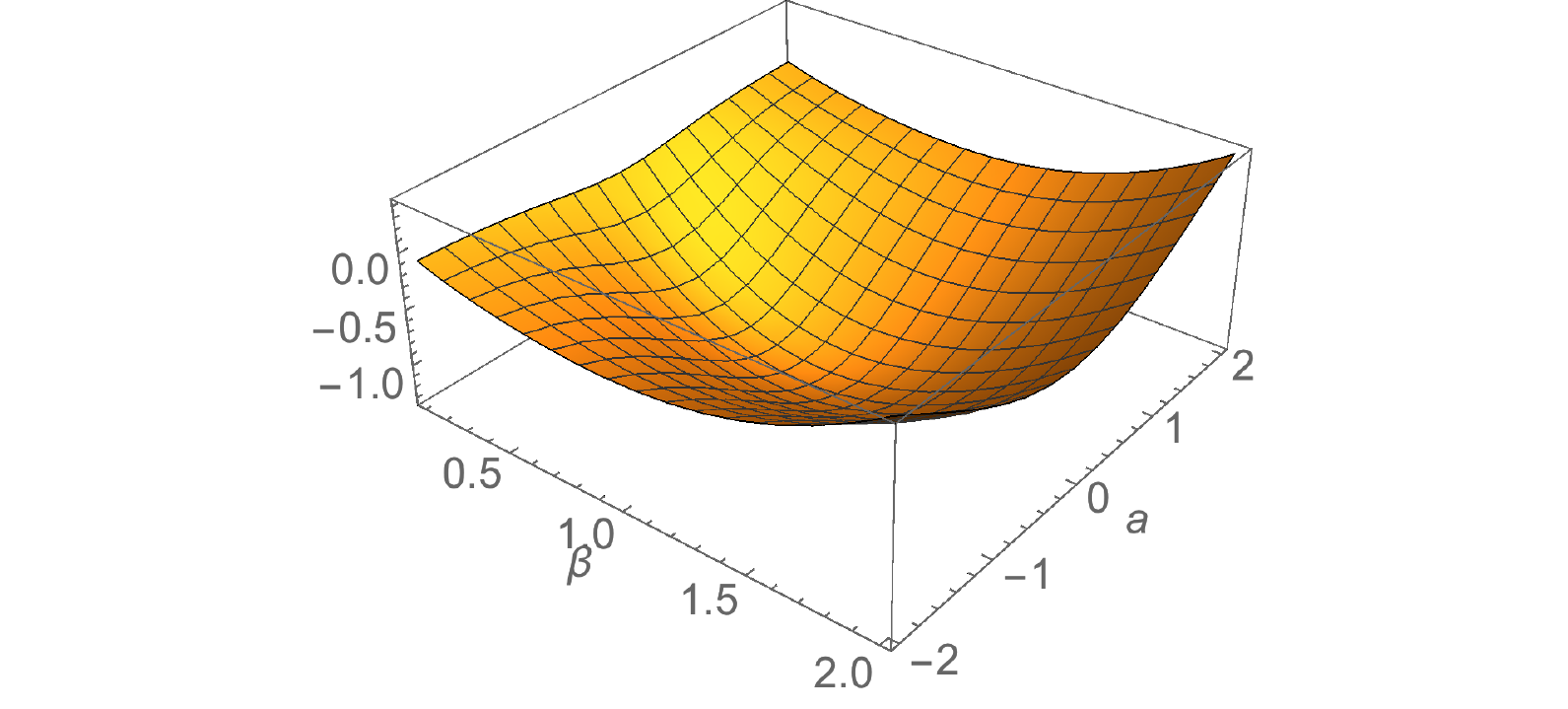} }
  \subfigure[\ Case II, repulsive potential.]
  { \label{f:Fig3Db} \includegraphics[width=0.95\columnwidth]{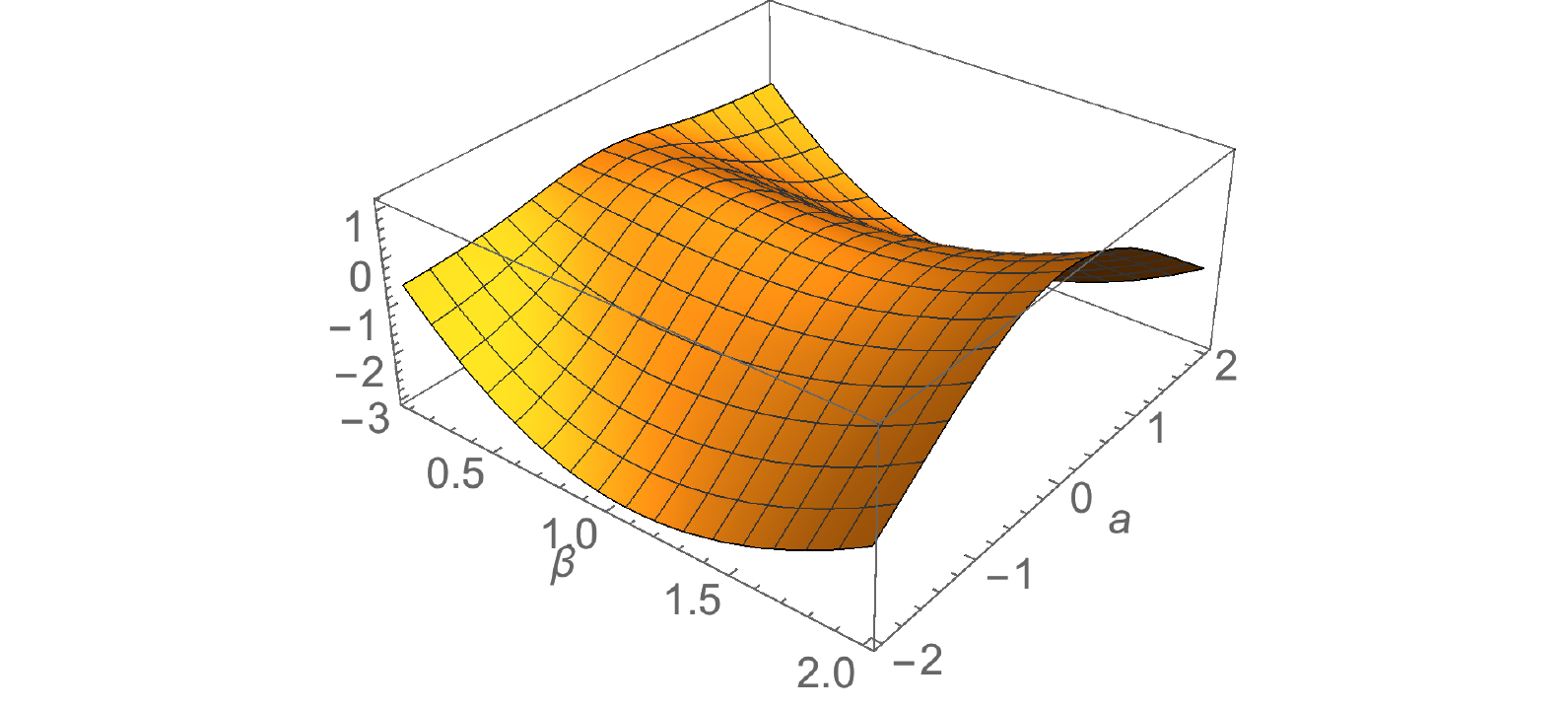} }
  \caption{\label{f:Fig3D}The 3D landscape $H(\beta,a)$ for $g=1$, $\tb^2=1$,
  and $\gamma = 1$.}
\end{figure*}
%
%

Using Derrick's theorem we explored whether the solution was a maximum or minimum of the energy landscape as a function of the stretching parameter $\beta$.
Here we would like to do a similar analysis to study whether the energy increases or decreases as we let $x \rightarrow x+a$ where $a$ is a small translation.  Again
we will posit that there is a translational instability if the energy $H[a]$ decreases as $a$ departs from zero.  Let us again consider NLSE plus real potential 
\begin{equation}\label{e:TS-1}
   \rmi \psi_t + \psi_{xx} + g|\psi|^{2\kappa} \psi - V(x) \psi= 0 \>,
\end{equation}
where $V(x) = -\lambda \tilde{b}^2 \sech^2(x)$. We want to see how the energy
of the system changes under the translation $x \rightarrow x+a $ with the normalization 
fixed by the requirement that the mass $M$ is preserved. Clearly choosing
\begin{equation}
   \psi_a = \psi(x+a)
\end{equation}
preserves the mass for the wave function of the exact solution whose $x$ dependence
is displayed by letting
\begin{equation}
\psi(x) = A(b, \gamma) \sech^\gamma(x) e^{i \gamma^2 t}
\end{equation}
It is also clear that both $H_1$ and $H_2$ remain unchanged under $x \rightarrow x + a$.  Only $H_3$ is not translationally invariant.  We define 
\begin{align}\label{e:TS-3}
   H_3(a, \gamma) 
   &= 
   \tint \dd{x} V(x)|\psi_{a}|^2
   \\
   &=
   \lambda \tilde{b}^2 A_{0}^{2} \int dx \sech^2 (x) \sech^{2\gamma}(x+a)
   \notag \\
   &=
   \lambda \tilde{b}^2 A_{0}^{2} \int dy \sech^2 (y-a) \sech^{2\gamma} (y) \>.
   \notag
\end{align}
We need to ensure that as a function of $a$ the energy is stationary.  Since both $H_1$ and $H_2$ are independent of $a$ we only need to consider:
\begin{equation}\label{e:TS-4}
   \pdv{H_3(a,\lambda)}{a} \Big |_{a=0}
\end{equation}
at $\beta =1$ and then calculate the second derivative at $\beta =1$.  Clearly 
\begin{align}\label{e:TS-4}
   &\pdv{H_{\beta}}{a} 
   = 
   \pdv{H_{3 a}}(a)
   = \\
   & 
   -2\lambda \tilde{b}^2 A_{0}^{2} 
   \tint \dd{y} \sech^2 (y-a) \tanh (y-a) 
   \sech^{2\gamma} (y) \>,
   \notag
\end{align}
which is indeed zero when evaluated at $a=0$, being an odd function of $y$.
The second derivative gives
\begin{align}\label{e:TS-4}
   &\pdv[2]{H_{a}}(a) 
   =
   \pdv[2]{H_{3 a}}{a}
   = \\ 
   &
   2\lambda \tilde{b}^2 A_{0}^{2} 
   \tint \dd{y} \sech^{2 \gamma}(y) \,
   [\, 3\sech^{4}(y-a) - 2\sech^{2} (y-a) \,] . 
   \notag
\end{align}
Evaluating this at $a=0$ we find: 
\begin{equation}
   \pdv[2]{H_{a}}{a} \Big |_{a=0}
   =
   2 \lambda \tilde{b}^2 A_{0}^{2} 
   \frac{\sqrt{\pi } \, \gamma \, \GammaF{\gamma +1}}
        {\GammaF{ \gamma + 5/2} } \>.
\end{equation}
Thus we indeed find that the solitary wave has translational instability if
$\lambda < 0$, i.e. for the repulsive potential, while it is stable in case
$\lambda > 0$, i.e. attractive potential. Note that the answer does not depend
on the sign of $g$.  We show in Fig.~\ref{f:Fig3D} the three-dimensional landscape for $H(\beta,a)$ for a stretching and displacement shift, $x \rightarrow \beta x + a$.  One can see stability for an attractive potential (case I) but a saddle point for a repulsive potential (case II).

%
%
\section{\label{s:Linear} Linear Stability Analysis}

The traditional way to study stability under small perturbations is to perform a linear stability analysis which we now present.  The results obtained agree with the
simpler analysis using Derrick's theorem and looking at the effects of translation on the energy landscape.  Starting with the NLSE equation in an external potential:
\begin{equation}\label{e:NLSE-1}
	\{\,
	\rmi \, \partial_t
	+
	\partial_{x}^2
	+
	g \, \abs{\psi(x,t)}^{2 \kappa}
	-
	V(x) \,
	\} \, \psi(x,t)
	=
	0 \>,
\end{equation}
which has solitary wave solutions $\psi(x,t)=\phi_{\omega}(x)\, \rme^{-\rmi\omega t}$, with $\phi_{\omega}(x) \in \mathbb{R}$.  Here $\phi_{\omega}(x)$ satisfies
\begin{equation}\label{e:NLSE-phi}
	\{\,
	\omega
	+
	\partial_{x}^2
	+
	g \, \abs{\phi_{\omega}(x)}^{2 \kappa}
	-
	V(x) \,
	\} \, \phi_{\omega}(x)
	=
	0 \> . 
\end{equation}
For $\omega = - \gamma^2$, one has the explicit expression:
\begin{equation}\label{e:NOSE-phi-sol}
   \phi_{-\gamma^2}(x)
   =
   A_0(\tb,\gamma) \, \sech^{\gamma}(x) \>.
\end{equation}
We consider perturbations in the form $\psi(x,t)=(\phi_{\omega}(x)+r(x,t)) \rme^{-\rmi\omega t}$ with $r(x,t)\in \mathbb{C}$ and linearize the equation (\ref{e:NLSE-1}) with respect to $r(x,t)$. The linearized equation is of the form
\begin{align}\label{e:LNLSE}
   &\{\,
      \omega 
      + 
      \rmi \partial_t 
      + 
      \partial_x^2+V(x)
      +
      g \abs{\phi_{\omega}(x)}^{2\kappa} \,
      \} \, r(x,t)
   \\
   & \qquad
   +
   2\kappa \, g \, \abs{\phi_{\omega}(x)}^{2\kappa} \Re{r(x,t)} = 0 \>.
   \notag
\end{align}
Because of the term $\Re{r(x,t)}$, Eq.~\ef{e:LNLSE} is not a $\mathbb{C}$-linear operator.
For computational convenience we separate the real and imaginary parts of $r(x,t)$ and define 
\begin{equation}
   R(x,t)
   =
   \begin{pmatrix}
      p(x,t)\\
      q(x,t)
   \end{pmatrix}
   =
   \begin{pmatrix}
      \Re{r(x,t)}\\
      \Im{r(x,t)}
   \end{pmatrix} \>.
\end{equation}
The equation \ef{e:LNLSE} can then be written as
\begin{equation}\label{e:MLNLSE}
   \partial_t R = \mathbf{JL} \, R
\end{equation}
where 
\begin{equation}
   \mathbf{J}
   =
   \begin{pmatrix}
      0 & 1 \\
      -1 & 0
   \end{pmatrix}
   \qc
   \mathbf{L}
   =
   \begin{pmatrix}
      L_+ & 0 \\
      0 & L_-
   \end{pmatrix}
\end{equation}
with self-adjoint operators
\begin{subequations}\label{e:SA}
\begin{align}
   L_{-}(\omega)
   &=
   -\partial_x^2 + V(x) - \omega - g\abs{\phi_{\omega}(x)}^{2\kappa} \>,
   \label{e:SAa} \\
   L_{+}(\omega)
   &=
   L_{-}(\omega) - 2\kappa g\abs{\phi_{\omega}(x)}^{2\kappa} \>.
   \label{e:SAb}
\end{align}
\end{subequations}
From \ef{e:NLSE-phi} and its derivative with respect to $\omega$, we find
\begin{equation}\label{e:SA-add}
   L_{-}(\omega) \, \phi_{\omega}(x)
   =
   0
   \qc
   L_{+}(\omega) \, \partial_{\omega} \phi_{\omega}(x)
   =
   \phi_{\omega}(x) \>.
\end{equation}
To explore the linear stability we consider eigenvalues for the operator $\mathbf{JL}$. 
Since 
\begin{equation}
   \mathbf{(JL)}^2
   =
   \begin{pmatrix}
      -L_{-}(\omega) L_{+}(\omega) & 0\\
      0 & -L_{+}(\omega)L_{-}(\omega)
    \end{pmatrix}
\end{equation}
and the nonzero eigenvalues of $(-L_{-} L_{+})$ and $(-L_{+}L_{-})$ coincide,
we can consider eigenvalues of the operator $(-L_-L_+)$ instead. If $(-L_-L_+)$ has eigenvalues with positive real part, so does $\mathbf{JL}$ and the solitary wave solutions are linearly unstable; otherwise, $\mathbf{JL}$ only has purely imaginary eigenvalues and the solitary wave solutions are spectrally stable.

For case (II) $g=+1$, $\lambda=-1$ we have $\omega=-\gamma^2$. The amplitude $\phi_{\omega}(x)$ is in $L^2$ and $L_-(\omega)\phi_{\omega}=0$, so $\phi_{\omega}$ is an eigenfunction of $L_-(\omega)$, corresponding to zero eigenvalue. Since $\phi_{\omega}$ is positive, 
 $L_-(\omega)$ is nonnegative and the kernel $\ker(L_-({\omega}))$ is span$\{\phi_{\omega}\}$ by  Proposition 2.8 in 
 \cite{doi:10.1137/0516034}. Due to (\ref{e:SA-add})  $\partial_{\omega}\phi_{\omega}$ is an eigenfunction of $L_-L_+$,  corresponding to zero eigenvalue. We will show that the smallest eigenvalue of $L_-L_+$ is negative.   According to Ref.~\cite{r:Sulem-Catherine-and-Sulem-Pierre-Louis:1999fk} the smallest eigenvalue of $L_-L_+$  is given by,
\begin{equation}
   \min \sigma_d(L_-L_+)
   =
   \min \Big \{\,
       \frac{\expval{ u, L_ + u }}
            {\expval{u, L_-^{-1} u }}, u\in \ker(L_-)^{\perp}
        \Big \} \> .
\end{equation}
$L_-$ is positive definite in $\ker(L_-)^{\perp}$, so the sign of smallest eigenvalue is decided by that of $\langle u, L_+u\rangle$. Since $\phi_{\omega}(x)$ is an even function, $\partial_{x}\phi_{\omega}(x)$ is an odd function in $\ker(L_-)^{\perp}$.
Due to the identity
\begin{equation}
  \partial_x (L_-\phi_{\omega}(x))
   =
   L_ + \partial_{x}\phi_{\omega}(x)+V'(x)\phi_{\omega}(x) = 0 \>,
\end{equation}
one has that $L_+\partial_{x}\phi_{\omega}(x)= -V'(x)\phi_{\omega}(x)$.  Hence we have 
\begin{align}
   &\expval{ \partial_{x}\phi_{\omega}, L_ + \partial_{x}\phi_{\omega} }
   =
   \expval{ \partial_{x}\phi_{\omega}, -V'(x)\phi_{\omega} }
   \\
   & \qquad\qquad
   =
   \frac{1}{2}\,
   \expval{ \phi_{\omega}, V''(x) \phi_{\omega} }
   \approx
   - 0.455 < 0 \>.
   \notag
\end{align}
It follows that $\expval{u, L_+u}$ can be negative and thus $(-L_-L_+)$ has at least one positive eigenvalue.  We conclude that the solitary wave solutions are linearly unstable.

For case (III) $g=-1$, $\lambda=+1$ we know that $L_-$ is nonnegative as in case (II).  Since $-2\kappa g\abs{\phi_{\omega}(x)}^{2\kappa}$ is positive, $L_+$ is positive as well. It implies that $-L_-L_+$ has only negative eigenvalues and the solitary wave solutions are spectrally stable.

%
%
\section{\label{s:fourtrial}Four parameter time-dependent trial wave function}

In order to study the dynamics in a collective coordinate approximation as well as determine the frequency of small oscillations (or the intitial growth of instabilities)
 we will now consider a four-parameter trial wave function of the form:
\begin{equation}\label{e:T-1}
   \tpsi(x,t)
   =
   A(t) \, \sech^{\gamma}[\, \beta(t) \, y(x,t) \,] \, 
   \rme^{\rmi \, \tphi(x,t)} \>,
\end{equation}
where
\begin{equation}\label{e:T-2}
   \tphi(x,t)
   =
   - 
   \theta(t)
   +
   p(t) \, y(x,t)
   +
   \Lambda(t) \, y^2(x,t) \>.
\end{equation}
Here we have put $y(x,t) = x - q(t)$. The parameter $\Lambda$ is related to the canonical conjugate variable to the average value  of
$y^2$.  It arises naturally in the Hartree-Fock approximation to the dynamics of the \Schrodinger\ equation \cite{ PhysRevD.34.3831}.

 It will be useful to define a reciprocal width parameter $G(t) = 1/\beta(t)$ and use this parameter as a generalized coordinate.  
Conservation of probability gives the ``mass'' equation,
\begin{equation}\label{e:T-3}
   M
   =
   \tint \dd{x} \rho(x,t)
   =
   G(t) \, A^2(t) \,  c_1[\gamma] , 
\end{equation}
where $c_1[\gamma]$ is given in Eq.~\ef{e:NLSE-9}.
In order to maintain probability conservation, we want to keep $M$ constant.  That is, we put
\begin{equation}\label{e:T-5}
   A^2(t)
   =
   \frac{M}{G(t) \, c_1[\gamma]}
\end{equation}
so $A(t)$ and $G(t)$ are not independent variables.  The phase $\theta(t)$ does not enter into the Hamiltonian, and we will ignore it in what follows.  The trial wave function we will assume is:
\begin{equation}\label{e:T-6}
   \tpsi(x,t)
   =
   \sqrt{\frac{M}{G \, c_1[\gamma]}}
    \, \sech^{\gamma}(y/G) \, 
   \rme^{\rmi \, [\, p \, y + \Lambda \, y^2 \,]} \>.
\end{equation}
The four variational parameters are labeled by
\begin{equation}\label{e:T-7}
   Q^i(t)
   =
   \qty{\, q(t), p(t), G(t), \Lambda(t) \,} \>. 
\end{equation}
Taking the appropriate derivatives we find that
\begin{equation}\label{e:T-11}
   L_0
   =
   \frac{\rmi}{2} \, \tint \dd{x}
   [\,
      \tpsi^{\ast} \, \tpsi_t
      -
      \tpsi^{\ast}_t \, \tpsi \,
   ]
   =
   \pi_i(Q) \, \dot{Q}^i \>,
\end{equation}
where
\begin{subequations}\label{e:T-12}
\begin{align}
   \pi_q
   &=
   p \, M \>,
   \label{e:T-12a} \\
   \pi_p
   &=
   0
   \label{e:T-12b} \\
   \pi_{\beta}
   &=
   0 \>,
   \label{e:T-12c} \\
   \pi_{\Lambda}
   &=
   - M G^2 \, c_2[\gamma] / c_1[\gamma]  \>,
   \label{e:T-12d}
\end{align}
\end{subequations}
with $c_2[\gamma]$ given in Eq.~\ef{e:DT-10.1}.
The only non-zero derivatives of the $\pi_i$ are
\begin{equation}\label{e:T-14}
   \partial_p \pi_q = M 
   \qc
   \partial_G \pi_{\Lambda}
   =
   - 2 M G \, c_2[\gamma] / c_1[\gamma] \>,
\end{equation}
so the symplectic tensor is
\begin{equation}\label{e:T-15}
   f_{ij}(Q)
   =
   M
   \begin{pmatrix}
      0 & -1 & 0 & 0 \\
      1 & 0 & 0 & 0 \\
      0 & 0 & 0 & -C \\
      0 & 0 & C & 0
   \end{pmatrix}
   \qc
   C = 2 \, G \, \frac{c_2[\gamma]}{c_1[\gamma]} \>,
\end{equation}
and the inverse is
\begin{equation}\label{e:T-16}
   f^{ij}(Q)
   =
   \frac{1}{M  C}
   \begin{pmatrix}
      0 & C & 0 & 0 \\
      -C & 0 & 0 & 0 \\
      0 & 0 & 0 & 1 \\
      0 & 0 & -1 & \\
   \end{pmatrix} \>.
\end{equation}
We also find that for the four-parameter trial wave function
\begin{equation}\label{e:T-17}
   H = H_1 + H_2 + H_3 \>,
\end{equation}
where
\begin{subequations}\label{e:T-18}
\begin{align}
   H_1
   &=
   \tint \dd{x} | \tpsi_x |^2
   \label{e:T-18a} \\
   &=
   M \, p^2
   +
   \frac{M}{G^2} \, \frac{\gamma}{2} \, \frac{c_1[\gamma+1]}{c_1[\gamma]}
   +
   4 \, M G^2 \Lambda^2 \, \frac{c_2[\gamma]}{c_1[\gamma]} \>,   
   \notag  \\
   H_2
   &=
   - \frac{g}{\kappa + 1} 
   \tint \dd{x} | \tpsi |^{2\kappa+2} 
   \label{e:T-18b} \\
   &=
   - 
   \frac{g M \, \gamma}{\gamma+1} \,
   \qty(\, \frac{M}{G \,c_1[\gamma]} \,)^{1/\gamma} \,
   \frac{c_1[\gamma+1]}{c_1[\gamma]}
   \notag \\
   H_3
   &=
   \tint \dd{x} V(x) \, | \tilde{\psi} |^2
   =
   - \lambda \, \tb^2  \, M \, 
   \frac{f_1[G,q,\gamma]}{c_1[\gamma]} \>,
   \label{e:T-18c}
\end{align}
\end{subequations}
with
\begin{equation}\label{e:T-19}
   f_1[G,q,\gamma]
   =
   \tint \dd{z} \sech^{2\gamma}(z) \sech^2( G z + q ) . 
\end{equation}
The Hamiltonian then becomes
\begin{align}\label{e:T-20}
   &\frac{H(Q)}{M}
   =
   p^2
   +
   \frac{\gamma}{2 \, G^2} \, \frac{c_1[\gamma+1]}{c_1[\gamma]}
   +
   4 \, G^2 \Lambda^2 \, \frac{c_2[\gamma]}{c_1[\gamma]}
   \\
   & \quad
   - 
   \frac{g \, \gamma}{\gamma+1} \,
   \qty(\, \frac{M}{G \,c_1[\gamma]} \,)^{1/\gamma} \,
   \frac{c_1[\gamma+1]}{c_1[\gamma]}
   -
   \lambda \, \tb^2 \, \frac{f_1[G,q,\gamma]}{c_1[\gamma]} \>.
   \notag
\end{align}
Derivatives of the Hamiltonian, $v_i = \partial_i H(Q)$, are given by
\begin{subequations}\label{e:T-21}
\begin{align}
   \frac{v_q}{M}
   &=
   2 \lambda \, \tb^2 
   \frac{f_2[G,q,\gamma]}{c_1[\gamma]} \>,
   \label{e:T-21a} \\
   \frac{v_p}{M}
   &=
   2 \, p \>,
   \label{e:T-21b} \\
   \frac{v_{G}}{M}
   &=
   -
   \frac{\gamma}{G^3} \, \frac{c_1[\gamma+1]}{c_1[\gamma]}
   +
   8 \, G \Lambda^2 \, \frac{c_2[\gamma]}{c_1[\gamma]}
   \label{e:T-21c} \\
   & \!\!\!\!\!\!\!\!
   +
   \frac{g}{\gamma+1} \, \frac{1}{G} \,
   \qty(\, \frac{M}{G \,c_1[\gamma]} \,)^{1/\gamma} \,
   \frac{c_1[\gamma+1]}{c_1[\gamma]}
   +
   2 \, \lambda \, \tb^2 \, \frac{f_3[G,q,\gamma]}{c_1[\gamma]} \>,
   \notag \\
   \frac{v_{\Lambda}}{M}
   &=
   8 \, G^2 \Lambda \, \frac{c_2[\gamma]}{c_1[\gamma]} \>.
   \label{e:T-21d}
\end{align}
\end{subequations}
where
\begin{subequations}\label{e:T-22}
\begin{align}
   &f_2[G,q,\gamma]
   =
   - \frac{1}{2} \, \partial_q f_1[G,q,\gamma] 
   \label{e:T-22a} \\
   & \quad
   =
   \tint \dd{z} \sech^{2\gamma}(z) \sech^2( G z + q ) \tanh( G z + q ) \>,
   \notag \\
   &f_3[G,q,\gamma]
   =
   - \frac{1}{2} \, \partial_G f_1[G,q,\gamma] 
   \label{e:T-22b} \\
   & \quad
   =
   \tint \dd{z} z \, \sech^{2\gamma}(z) \sech^2( G z + q ) \tanh( G z + q) \> . 
   \notag
\end{align}
\end{subequations}
From Eq.~\ef{e:VT-10} and using \ef{e:T-16}, we find
\begin{subequations}\label{e:T-23}
\begin{align}
   \dot{q}
   &=
   \frac{v_p}{M}
   =
   2 \, p \>,
   \label{e:T-23a} \\
   \dot{p}
   &=
   - \frac{v_q}{M}
   =
   - 2 \lambda \, \tb^2 \, \frac{f_2[G,q,\gamma]}{c_1[\gamma]} \>,
   \label{e:T-23b} \\
   \dot{G}
   &=
   \frac{v_{\Lambda}}{M C} 
   =
   4 \, G \, \Lambda \>,
   \label{e:T-23c} \\
   \dot{\Lambda}
   &=
   - \frac{v_{G}}{M C}
   =
   - \frac{1}{2 G} \, \frac{c_1[\gamma]}{c_2[\gamma]} \, \frac{v_{G}}{M}  \notag \\
   &=
   -
   4 \Lambda^2
   +
   \frac{\gamma}{2 \, G^4} \, \frac{c_1[\gamma+1]}{c_2[\gamma]}
   \label{e:T-23d}  \\
   & \quad 
   -
   \frac{g}{2(\gamma+1)} \, \frac{1}{G^2} \,
   \qty(\, \frac{M}{G \,c_1[\gamma]} \,)^{1/\gamma} \,
   \frac{c_1[\gamma+1]}{c_2[\gamma]} \notag \\
   & \quad
   -
   \frac{\lambda \, \tb^2}{G} \,\frac{f_3[G,q,\gamma]}{c_2[\gamma]} \>.
   \notag
\end{align}
\end{subequations}
Recall that from \ef{e:T-5},
\begin{equation}\label{e:T-24}
   g \, \qty(\, \frac{M}{G(t) \, c_1[\gamma]} \,)^{1/\gamma}
   =
   g \, A^{2/\gamma}(t) \>,
\end{equation}
and since $G(0) = 1$ at $t=0$, we can use Eq.~\ef{e:NLSE-6} to find:
\begin{equation}\label{e:T-24.1}
   g \, \qty(\, \frac{M}{c_1[\gamma]} \,)^{1/\gamma}
   =
   g \, A^{2/\gamma}_0
   =
   \gamma ( \gamma + 1 ) - \lambda \tb^2 \>.
\end{equation}
So the dynamic equations \ef{e:T-23} reduce to:
\begin{subequations}\label{e:T-25}
\begin{align}
   \dot{q}
   &=
   2 \, p \>,
   \label{e:T-25a} \\
   \dot{p}
   &=
   - 2 \lambda \, \tb^2 \, \frac{f_2[G,q,\gamma]}{c_1[\gamma]} \>,
   \label{e:T-25b} \\
   \dot{G}
   &=
   4 \, G \, \Lambda \>,
   \label{e:T-25c} \\
   \dot{\Lambda}
   &=
   -
   4 \, \Lambda^2 
   +
   \frac{1}{ G^4} \, \frac{ \gamma^2}{2\gamma+1} \,
   \frac{c_1[\gamma]}{c_2[\gamma]} 
   \label{e:T-25d} \\
   & \quad
   -
   \frac{\gamma \, [\, \gamma ( \gamma + 1 ) - \lambda \, \tb^2 \,]}
   { (2\gamma+1)(\gamma+1)} \, \frac{1}{G^{2+1/\gamma}} \,
   \frac{c_1[\gamma]}{c_2[\gamma]} \notag \\
   & \quad
   -
   \frac{\lambda \, \tb^2}{G} \,\frac{f_3[G,q,\gamma]}{c_2[\gamma]} \>,
   \notag
\end{align}
\end{subequations}
and are independent of $g$.  We need to remember that for case III when $g=-1$ and $\lambda=+1$ there is no solution unless $\tb^2 > \gamma (\gamma +1)$.

%
%
\subsection{\label{s:linear}Small oscillations}

We require that the variational wave function starts out so that it agrees with the exact solutions at $t=0$,
\begin{equation}\label{e:L-1}
   \tpsi(x,0)
   =
   \psi_0(x,0) \>,
\end{equation}
with $A_0$ fixed by \ef{e:T-24.1}.
This means that we want to choose
\begin{equation}\label{e:L-2}
   q_0 = 0
   \qc
   p_0 = 0
   \qc
   G_0 = 1
   \qc
   \Lambda_0 = 0 \>.
\end{equation}
Setting
\begin{align}\label{e:L-3}
   q(t)
   &=
   q_0 + \delta q(t)
   \qc
   p(t)
   =
   p_0 + \delta p(t)
   \qc   \notag \\
   G(t)
   &=
   G_0 + \delta G(t)
   \qc
   \Lambda(t)
   =
   \Lambda_0 + \delta \Lambda(t) \>,
\end{align}
to first order, we have
\begin{subequations}\label{e:L-4}
\begin{align}
   f_1[G,q,\gamma]
   &=
   f_1[1,0,\gamma] + g_1[\gamma] \, \delta G \>,
   \label{e:L-4a} \\
   f_2[G,q,\gamma]
   &=
   f_2[1,0,\gamma] + g_2[\gamma] \, \delta q \>,
   \label{e:L-4b} \\
   f_3[G,q,\gamma]
   &=
   f_3[1,0,\gamma] + g_3[\gamma] \, \delta G \>,  
   \label{e:L-4c}
\end{align}
\end{subequations}
where
\begin{subequations}\label{e:L-5}
\begin{align}
   f_1[1,0,\gamma]
   &=
   \tint \dd{z} \sech^{2\gamma + 2}(z)
   =
   c_1[\gamma+1] \>,
   \label{e:L-5a} \\
   f_2[1,0,\gamma]
   &=
   0 \>,
   \label{e:L-5b} \\
   f_3[1,0,\gamma]
   &=
   \tint \dd{z} z \, \sech^{2\gamma + 2}(z) \tanh(z) 
   \label{e:L-5c} \\
   &=
   \frac{c_1[\gamma+1]}{2(\gamma+1)} \>,
   \notag
\end{align}
\end{subequations}
where in the last term we have integrated by parts, and
\begin{subequations}\label{e:L-6}
\begin{align}
   g_1[\gamma]
   &=
   - 2 \tint \dd{z} z \, \sech^{2\gamma + 2}(z) \tanh(z) 
   \label{e:L-6a} \\
   &=
   - \frac{2 \gamma}{(2 \gamma + 1)( \gamma + 1) } \, c_1[\gamma] \>,
   \notag \\
   g_2[\gamma]
   &=
   \tint \dd{z} \sech^{2\gamma + 2}(z) \,
   [\, 1 - 3 \tanh^2(z) \,] \label{e:L-6b} \\
   &=
   \frac{4 \gamma^2 }{(2\gamma + 1)(2\gamma + 3)} \, c_1[\gamma] \>,
   \notag \\
   g_3[\gamma]
   &=
   \tint \dd{z} z^2 \, \sech^{2\gamma + 2}(z) \,
   [\, 1 - 3 \tanh^2(z) \,]
   \label{e:L-6c} \\
   &=
   - 2 \, c_2[\gamma+1] + 3 \, c_2[\gamma+2] \>.
   \notag
\end{align}
\end{subequations}
As a consistency check, at $t=0$ Eqs.~\ef{e:T-25} must all give values consistent with the initial conditions.  It is obvious that the first three equations are consistent, namely they all evaluate to zero.  At $t=0$, Eq.~\ef{e:T-25d} becomes
\begin{align}\label{e:L-8}
   &\frac{c_1[\gamma+1]}{c_2[\gamma]} \,
   \Bigl \{\,
      \frac{\gamma}{2}
      -
      \frac{1}{2(\gamma+1)} \,
      [\, \gamma ( \gamma + 1 ) - \lambda \, \tb^2 \,]
      \\
      & \qquad\qquad\qquad
      -
      \frac{\lambda \, \tb^2}{2(\gamma + 1)} \,
   \Bigr \}
   =
   0 \>,
   \notag
\end{align}
as required.  Substituting Eqs.~\ef{e:L-3} into \ef{e:T-23} gives
\begin{subequations}\label{e:L-9}
\begin{align}
   \delta \dot{q}
   &=
   2 \, \delta p \>,
   \label{e:L-9a} \\
   \delta \dot{p}
   &=
   - 
   \lambda \, \tb^2 \,
   \frac{8 \gamma (\gamma+1)}{(2\gamma + 1)(2\gamma + 3)} \, \delta q \>, 
   \label{e:L-9b} \\
   \delta \dot{G}
   &=
   4 \, \delta \Lambda \>,
   \label{e:L-9c} \\
   \delta \dot{\Lambda}
   &=
   \biggl \{\,
      -
      2 \gamma \, \frac{c_1[\gamma+1]}{c_2[\gamma]} 
      \label{e:L-9d} \\
      & \qquad  
      +
      \frac{2\gamma + 1}{2 \gamma(\gamma+1)} \,
      [\, \gamma ( \gamma + 1 ) - \lambda \, \tb^2 \,] \,
      \frac{c_1[\gamma+1]}{c_2[\gamma]}
      \notag \\
      & \qquad
      +
      \lambda \, \tb^2 \,\frac{f_3[1,0,\gamma]}{c_2[\gamma]}
      -
      \lambda \, \tb^2 \,\frac{g_3[\gamma]}{c_2[\gamma]}
   \biggr \} \, \delta G \>.
   \notag
\end{align}
\end{subequations}
So to first order, the $(q,p)$ and $(G,\Lambda)$ modes uncouple and reduce to equations of the form:
\begin{equation}\label{e:L-10}
   \delta \ddot{q} + \omega_q^2 \, \delta q = 0
   \qc
   \delta \ddot{G} + \omega_G^2 \, \delta G = 0 \>,
\end{equation}
where the $(q,p)$ mode frequency is given by
\begin{equation}\label{e:L-11}
   \omega_q^2
   =
   4 \lambda \, \tb^2 \, \frac{g_2[\gamma]}{c_1[\gamma]}
   =
   \lambda \, \tb^2 \,
   \frac{16 \gamma^2}{(2\gamma + 1)(2\gamma + 3)} \>,
\end{equation}
independent of $g$.  For $\lambda = +1$, translational motion is \emph{stable} for all $\gamma$, and \emph{unstable} for $\lambda = -1$.  This is easily explained by the fact that as long as the solitary wave is near $q=0$, for $\lambda = +1$ it sees an attractive force that brings it back to the origin.  For the opposite sign, $\lambda = -1$, it sees a repulsive force that moves it from the origin, assuming it maintains its shape.  We illustrate this behavior in Sec.~\ref{ss:trans} by a numerical solution of \Schrodinger's equation using a split-operator method. 

The $(G,\Lambda)$ mode frequency is given by
\begin{equation}\label{e:L-12}
   \omega_G^2
   =
   A(\gamma) + \lambda \, \tb^2 \, B(\gamma) \>.
\end{equation}
where
\begin{align}
   A(\gamma)
   &=
   \frac{ 4 \gamma ( 2 \gamma - 1) }{ 2 \gamma + 1 } \,
   \frac{c_1(\gamma)}{c_2[\gamma]} \>,
   \label{e:L-13} \\
   B(\gamma)
   &=
   4 \,
   \biggl \{\,
      \frac{1}{2 \gamma + 1} \, \frac{c_1(\gamma)}{c_2[\gamma]}
      -
      \frac{ 2 \, c_2[\gamma+1] - 3 \, c_2[\gamma+2] }{c_2[\gamma]} \,
   \biggr \} \>,
   \notag
\end{align}
and the critical value of $\tb^2$ is then given by
\begin{align}\label{e:L-14}
   \tb_{\text{crit}}^2
   &=
   - \lambda \, \frac{A(\gamma)}{B(\gamma)}
   \\
   &=
   - \lambda \,
   \frac{\gamma(2\gamma-1)}
   {1 
    - 
    (2\gamma+1) \,
    \frac{\displaystyle 2\, c_2[\gamma+1] - 3\, c_2[\gamma+2]}
         {\displaystyle c_1[\gamma]}  } \>,
   \notag
\end{align}
which is the same result for $\tb_{\text{crit}}^2$ that we found in Eq.~\ef{e:DT-12} using Derrick's theorem.  Here in addition, we find a value for the $(G,\Lambda)$ mode frequency $\omega_G^2$ in Eq.~\ef{e:L-12}.  Plots of $\tb_{\text{crit}}^2$, which are the same as Derrick's theorem, are shown in Figs.~\ref{f:Fig1}.

%
%
\subsection{\label{ss:dynamics}Dynamics of the Collective coordinates}

In this section we plot representative time evolutions of the collective coordinates of the four-parameter variational calculation given in Eqs.~\ef{e:T-25} for the three cases.  The parameters $(\,q, p, G, \Lambda \,)$ can be related to the expectation vales of $\langle \, x, p_{\text{op}}, x^2, x p_{\text{op}} \,\rangle$ where $p_{\text{op}} = - i \partial/\partial x$.  Here $\langle x \rangle /M = q(t)$, etc. 
One can show that using the equations obeyed by these four collective coordinates, that the evolution equations for these expectation values are \emph{exactly} satisfied.  We expect, and find that these time evolutions qualitatively agree with the numerically calculated values of these collective coordinates especially when the actual form of the numerically determined wave function is preserved.  In case III, although we predict the translational instability quite well, we did not anticipate that the form of the wave function would bifurcate.  Of course {\it a posteriori} one could assume a two humped variational wave function with more parameters to actually capture better the time evolution of the initial exact solution.  What we find, is that when we are in the oscillatory regime of either $q(t)$ or $G(t)$, the oscillation period determined from the small oscillation equations is in good agreement with what is found from the dynamical evolution of $q(t)$ and $G(t)$ from their evolution equations. 
 
%
%
\subsubsection{\label{glpp} $g=+1$ and $\lambda = +1$ (Case I)} 

\begin{figure*}[t]  
   \centering
   \subfigure[\ $q(t)$ \vs\ $t$]
   { \label{f:qt-case-Ia} 
     \includegraphics[width=0.9\columnwidth] {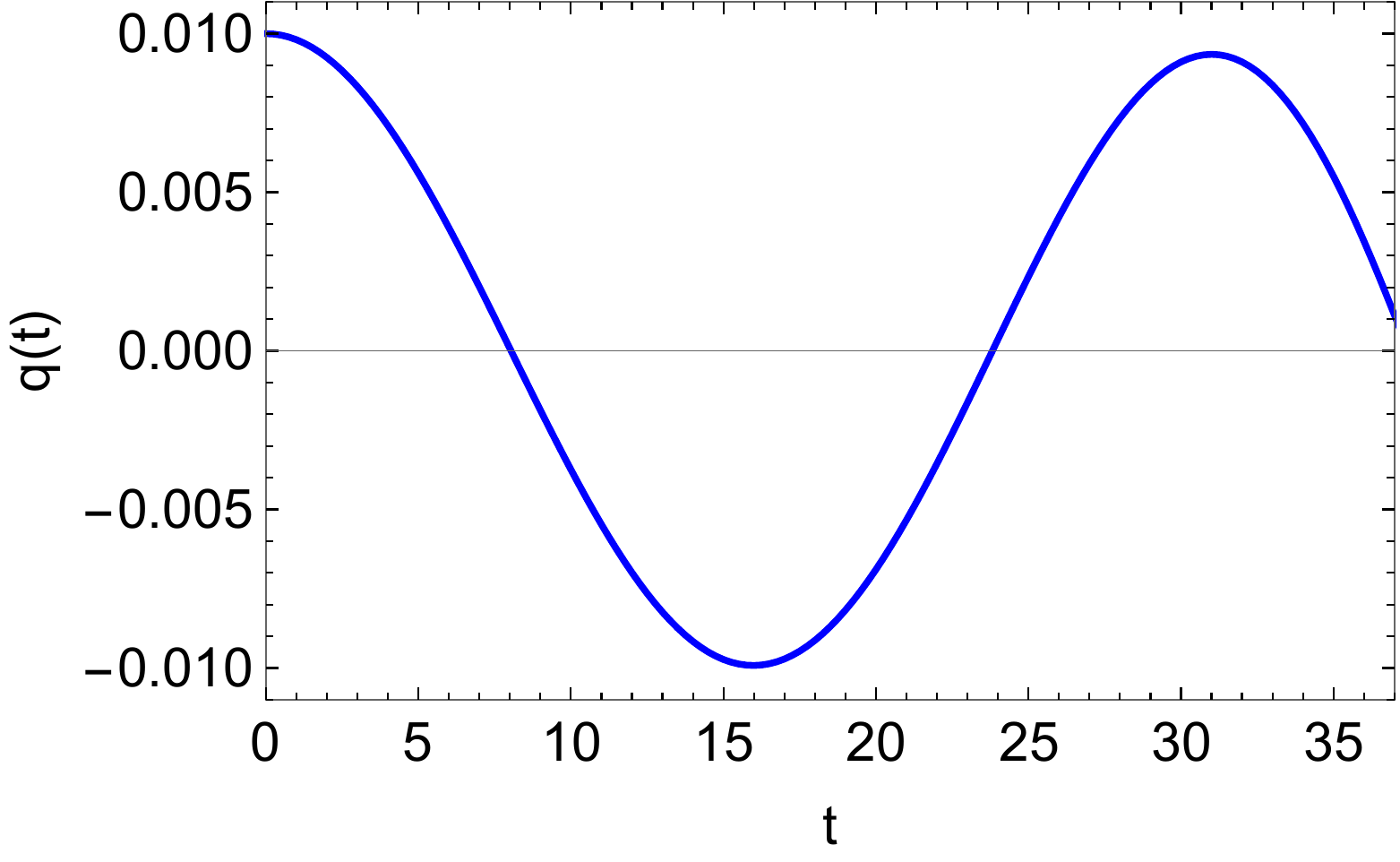} }
   \subfigure[\ $G(t)$ \vs\ $t$]
   { \label{f:Gt-case-Ia}  
     \includegraphics[width=0.9\columnwidth] {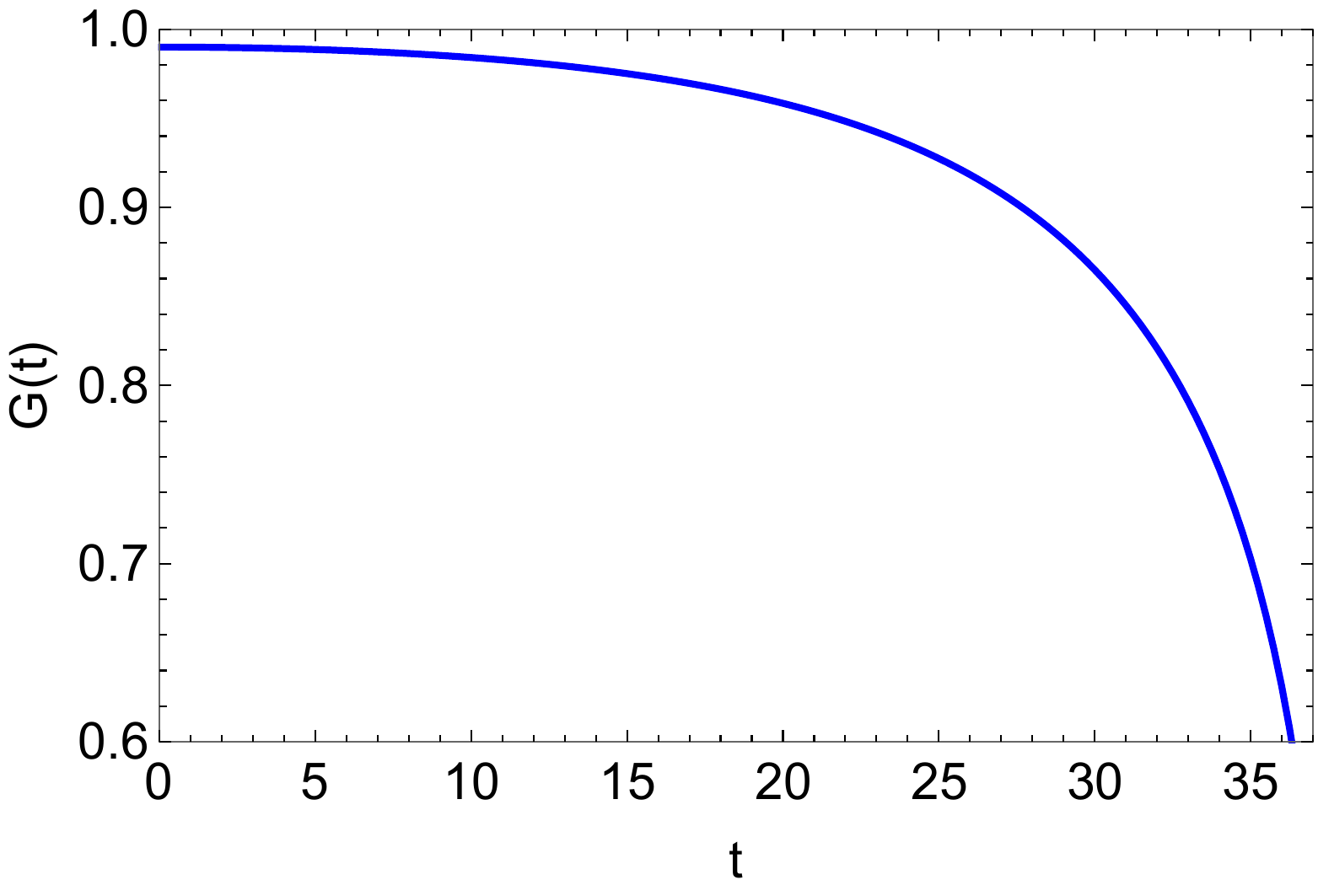} }
   \caption{\label{f:qGt-case-Ia} Case I for $\kappa=5/2$ and 
   $ \tb^2=0.1$.  Initial conditions are $G(0)=0.99$, $G'(0) = 0$,
   $q(0) = 0.01$, and $p(0) = 0$.}
\end{figure*}

\begin{figure*}[t]  
   \centering
   \subfigure[\ $q(t)$ \vs\ $t$]
   { \label{f:qt-case-Ib}
     \includegraphics[width=0.9\columnwidth] {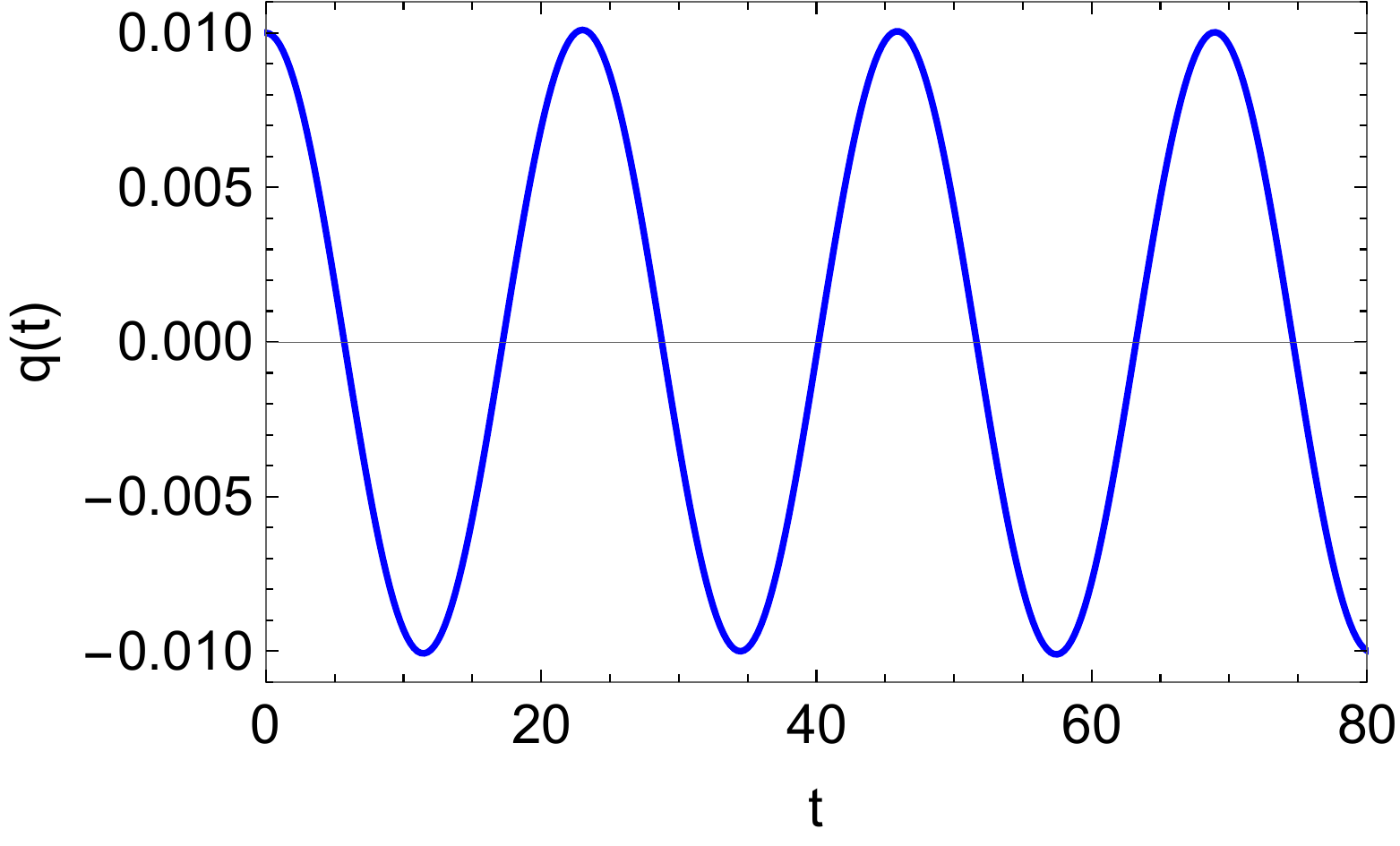} }
   \subfigure[\ $G(t)$ \vs\ $t$]
   { \label{f:Gt-case-Ib}
     \includegraphics[width=0.9\columnwidth] {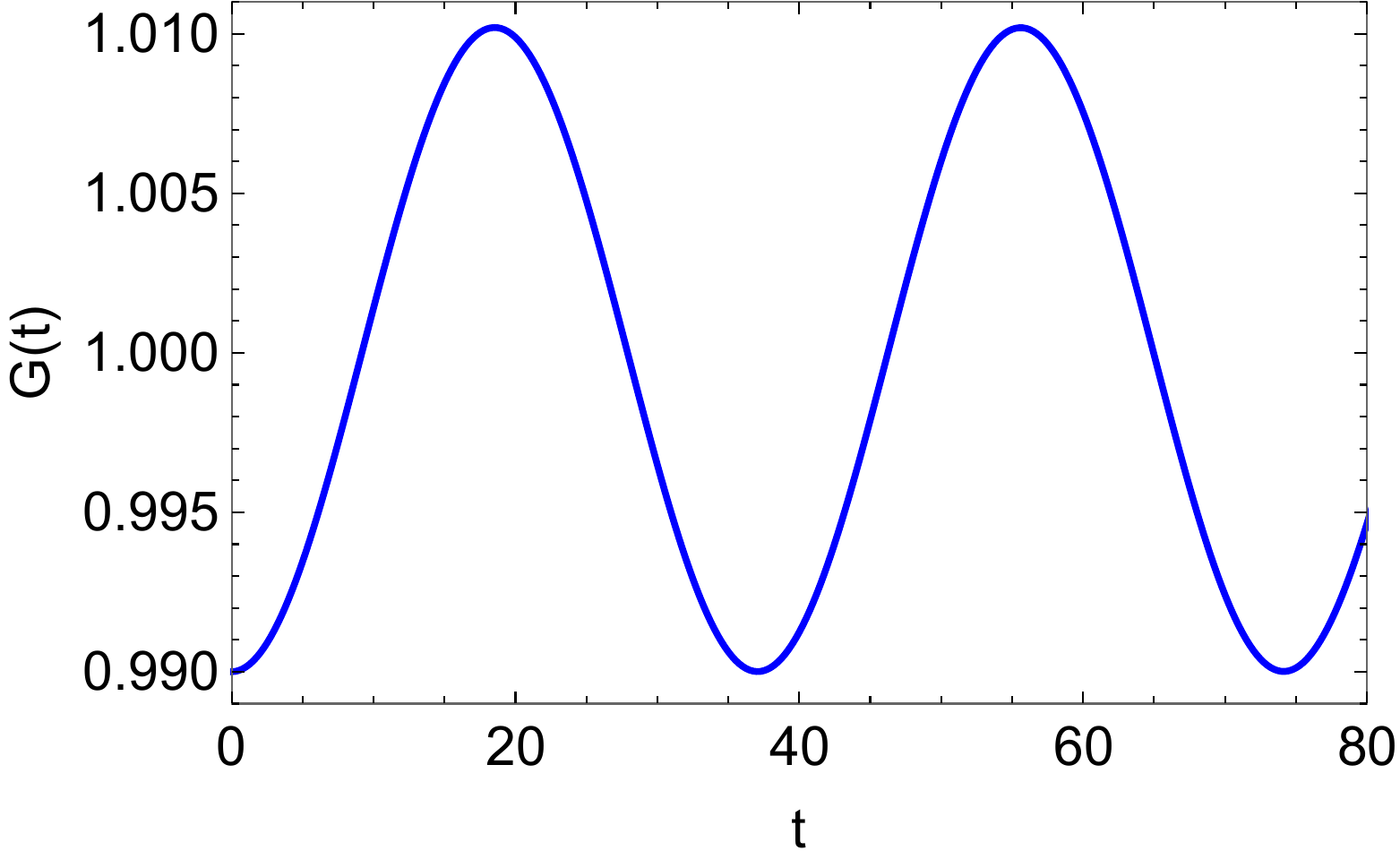} } 
   \caption{\label{f:qGt-case-Ib} Case I for $\kappa=5/2$ and 
   $\tb^2=0.2$. Initial conditions are $G(0)=0.99$, $G'(0) = 0$,
   $q(0) = 0.01$, and $p(0) = 0$.}
\end{figure*}

This is the case that we studied in our previous paper \cite{Cooper:2017aa}. However, in that paper we did not consider oscillations in the spatial direction, nor did we numerically solve the NLSE to compare with our analytic results.  For this case in the $G$ unstable 
region the behavior in $q$ is oscillatory in our approximation.  We will choose $\kappa = 5/2$ and two values of $\tb^2 =1/10$ which is in the unstable regime
and $\tb^2= 1/5$ which is in the stable regime to display the two types of behavior for the parameters $q(t)$ and $G(t)$ as a function of time.
For the blowup case, the period for $q(t)$ is oscillatory with a period of $T=32.5$, which agrees with the numerical results in Fig.~\ref{f:qt-case-Ia}.  However $G(t)$ blows up, $G(t) \rightarrow 0$, as shown in Fig.~\ref{f:Gt-case-Ia}.  
%
%
%
%
For $\tb^2 = 1/5$ one is in the oscillatory regime for $G(t)$ and $q(t)$ and we get the results shown in Figs.~\ref{f:qGt-case-Ib}.  For this case the small oscillation equation predicts that the period of $q(t)$ is $T=23$, and the period of $G(t)$ is $T=37$.
%
%
%
%

%
%
\subsubsection{\label{glpm} $g = +1$ and $\lambda = -1$ (Case II)}

\begin{figure*}[t]  
   \centering
   \subfigure[\ $q(t)$ \vs\ $t$]
   { \label{f:qt-case-IIa}
     \includegraphics[width=0.9\columnwidth]{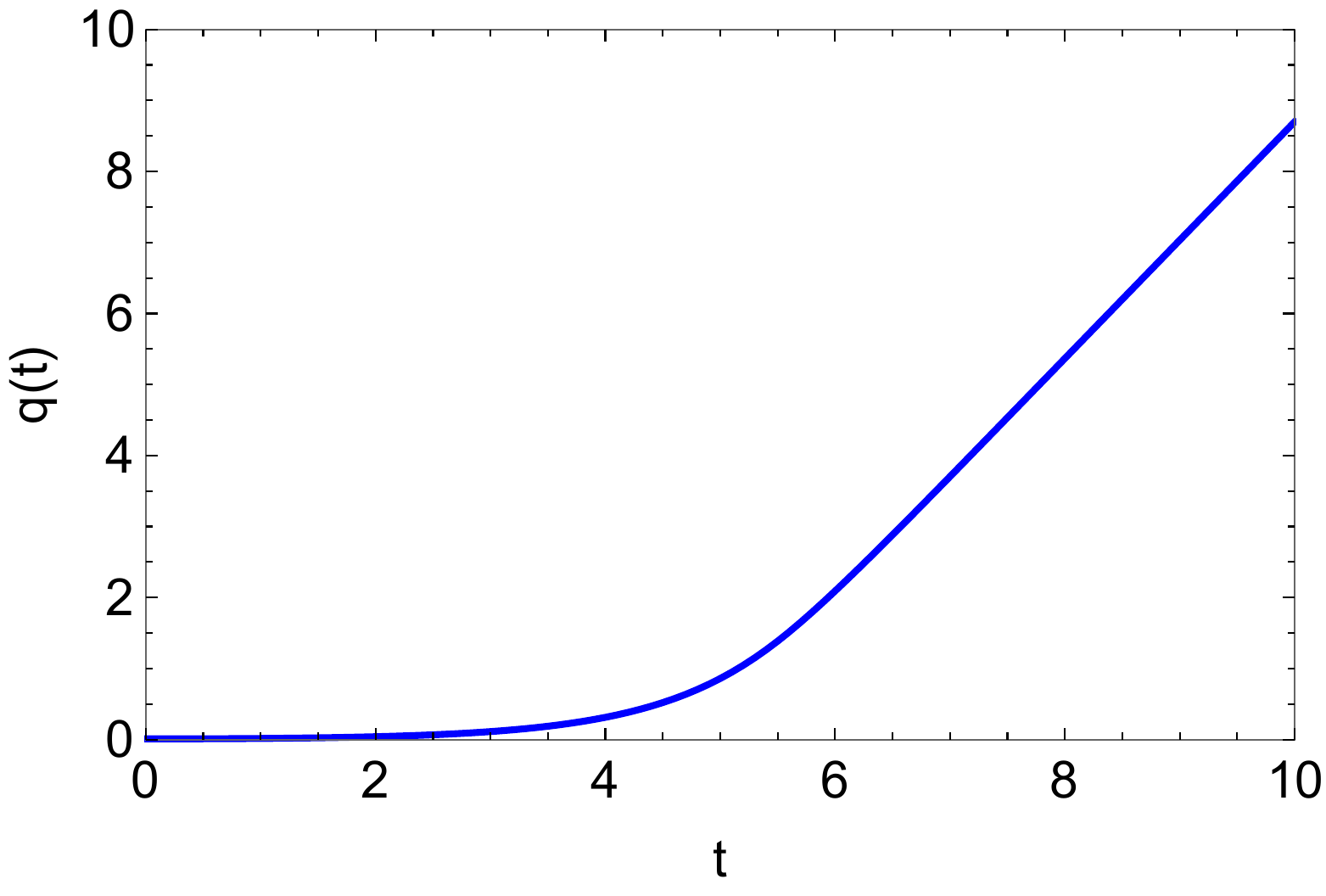} }
   \subfigure[\ $G(t)$ \vs\ $t$]
   { \label{f:Gt-case-IIa}
     \includegraphics[width=0.9\columnwidth]{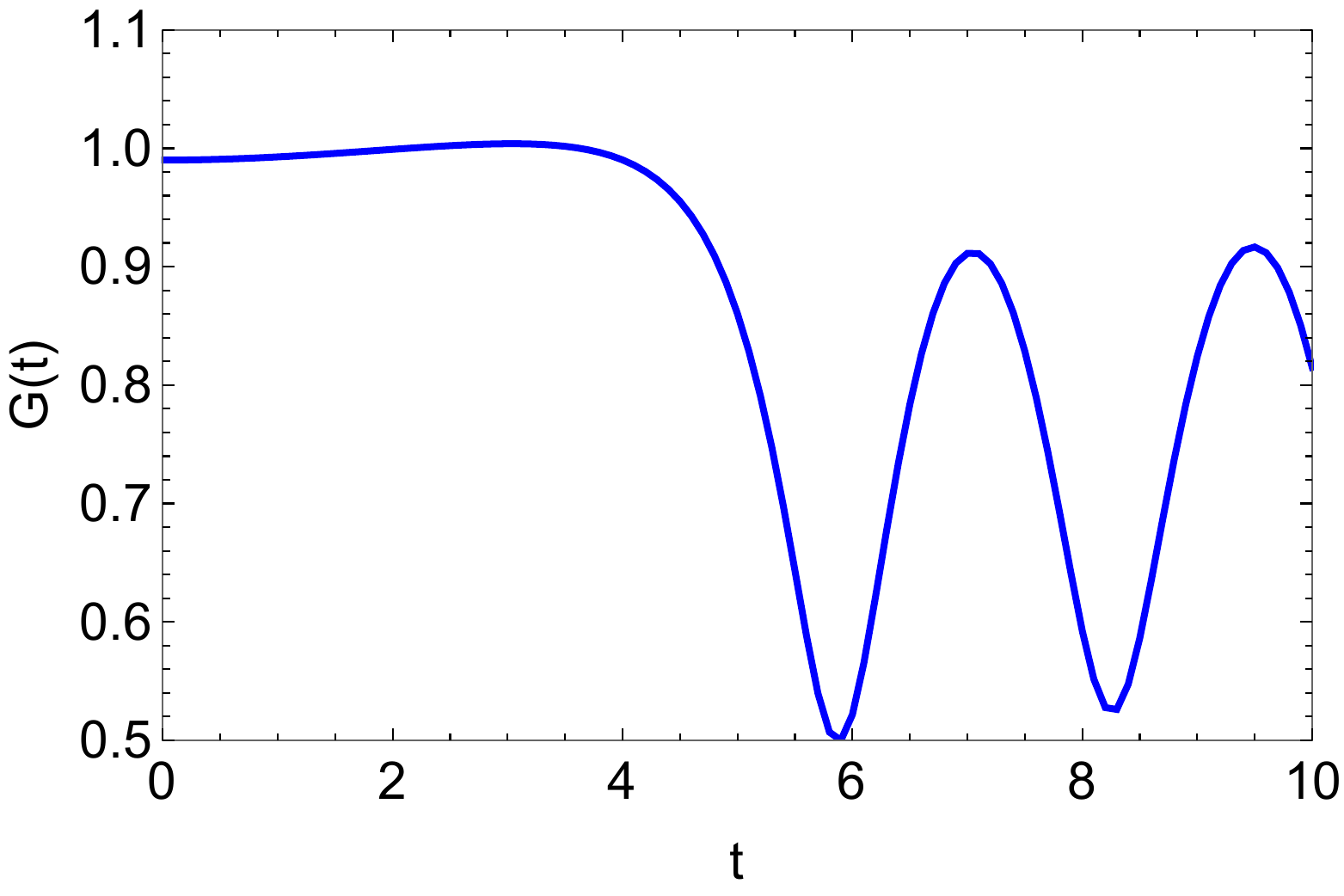} }
   \caption{\label{f:qGt-case-IIa} Case II for $\kappa=1$ and $\tb^2 = 1$.
   Initial conditions are $G(0)=0.99$, $G'(0) = 0$,
   $q(0) = 0.01$, and $p(0) = 0$.}
\end{figure*}

The case where we have an exact solution for a repulsive potential leads to the most unexpected behavior, as we will show later in our numerical simulations.
The first interesting thing is that although Derrick's theorem shows that the answer is stable to changing the width when $\kappa < 2$ and $\tb^2$ is below $\tb^2_{\text{c}}$ in Fig.~\ref{f:Fig1b}, we find that if we shift the position by a small amount, because of the repulsive potential, the solitary wave is pushed out of the region of the potential and then oscillates about a potential free solution in this approximation.  This result is suggested by the four-component variational calculation, and confirmed by a numerical calculation shown in Fig.~\ref{f:trans-b}.  Choosing $\kappa=1$ and $\tb^2 = 1$, which is in the regime which is stable to width changes, the  solutions of the dynamic equations \ef{e:T-24} give the results shown in Fig.~\ref{f:qGt-case-IIa}.
%
%
%
%


%
%
\begin{figure*}[t]  
\label{blowq}
   \centering
   \subfigure[\ $q(t)$ \vs\ $t$]
   { \label{f:qt-case-IIb}
     \includegraphics[width=0.9\columnwidth]{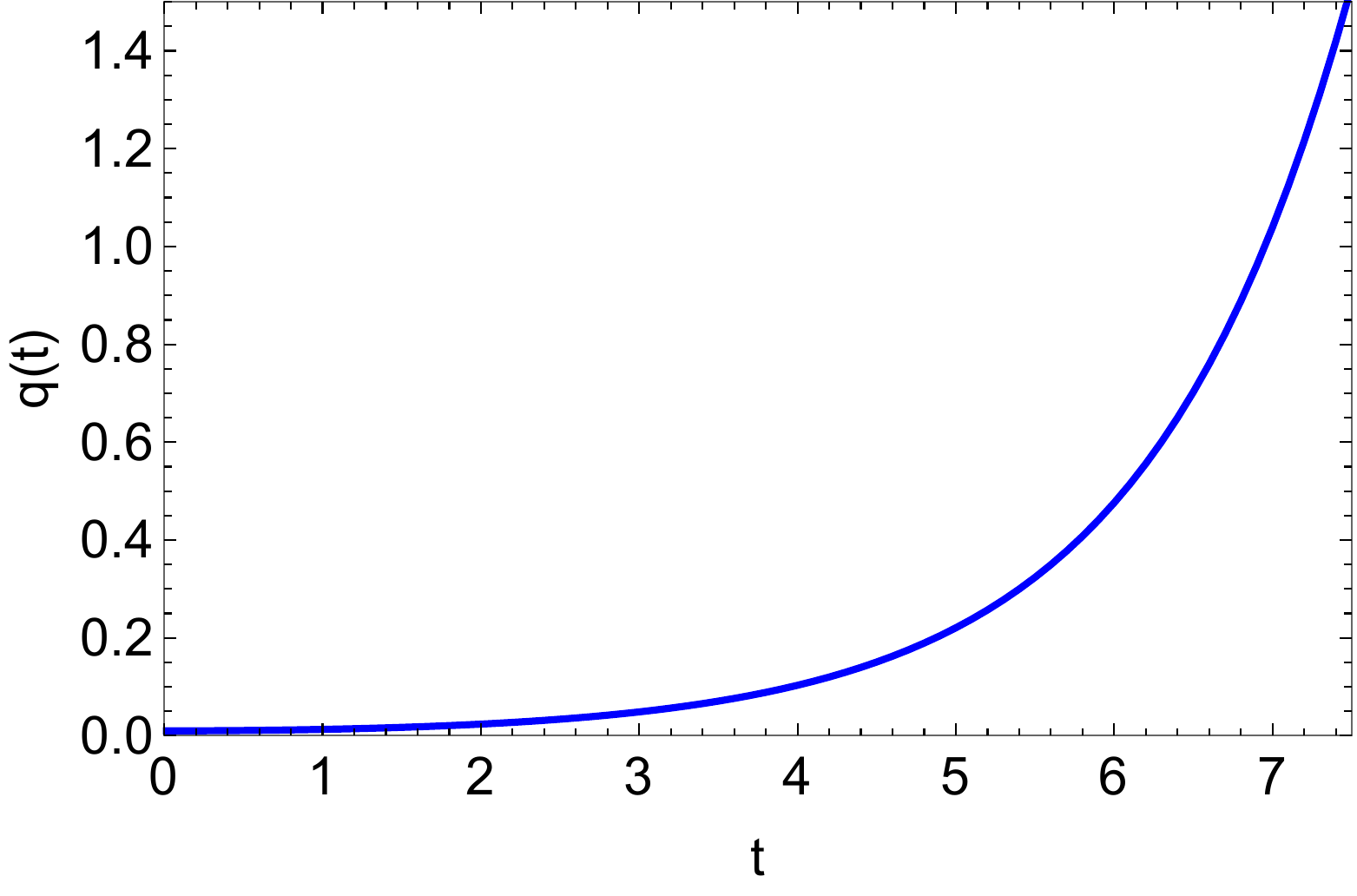} }
   \subfigure[\ $G(t)$ \vs\ $t$]
   { \label{f:Gt-case-IIb}
     \includegraphics[width=0.9\columnwidth]{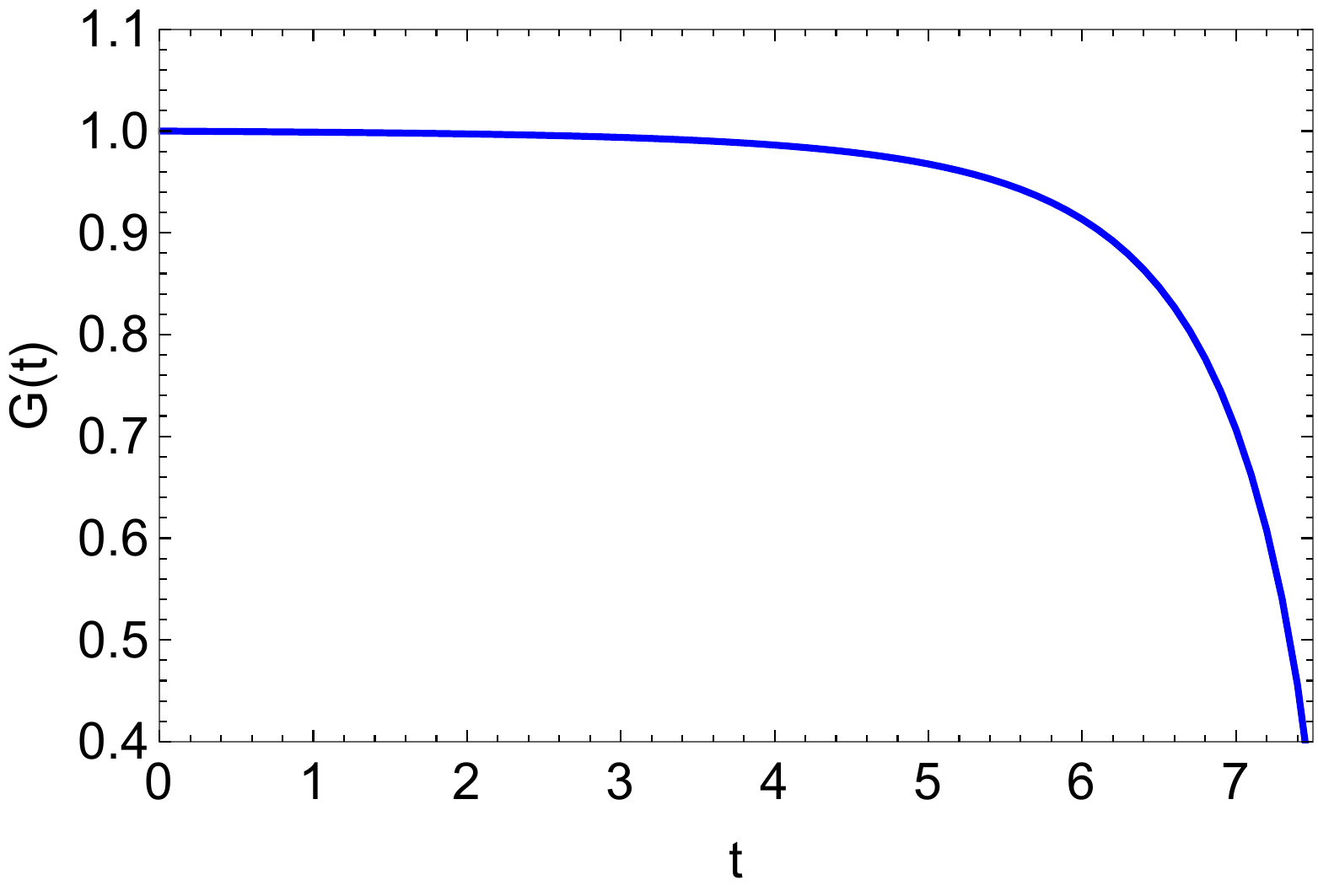} }
   \caption{\label{f:qGt-case-IIb} Case II for $\kappa=1.8$ and $\tb^2 = 1$.  
   Initial conditions are $G(0)=1.0$, $G'(0) = -0.001$, 
   $q(0) = 0.01$, and $p(0) = 0$.}
\end{figure*}
%
%

%
%
\subsubsection{\label{glmp} $g = -1$ and $\lambda = +1$ (Case III)} 

\begin{figure*}[t]  
   \centering
   \subfigure[\ $q(t)$ \vs\ $t$]
   { \label{f:qt-case-IIIa}
     \includegraphics[width=0.9\columnwidth]{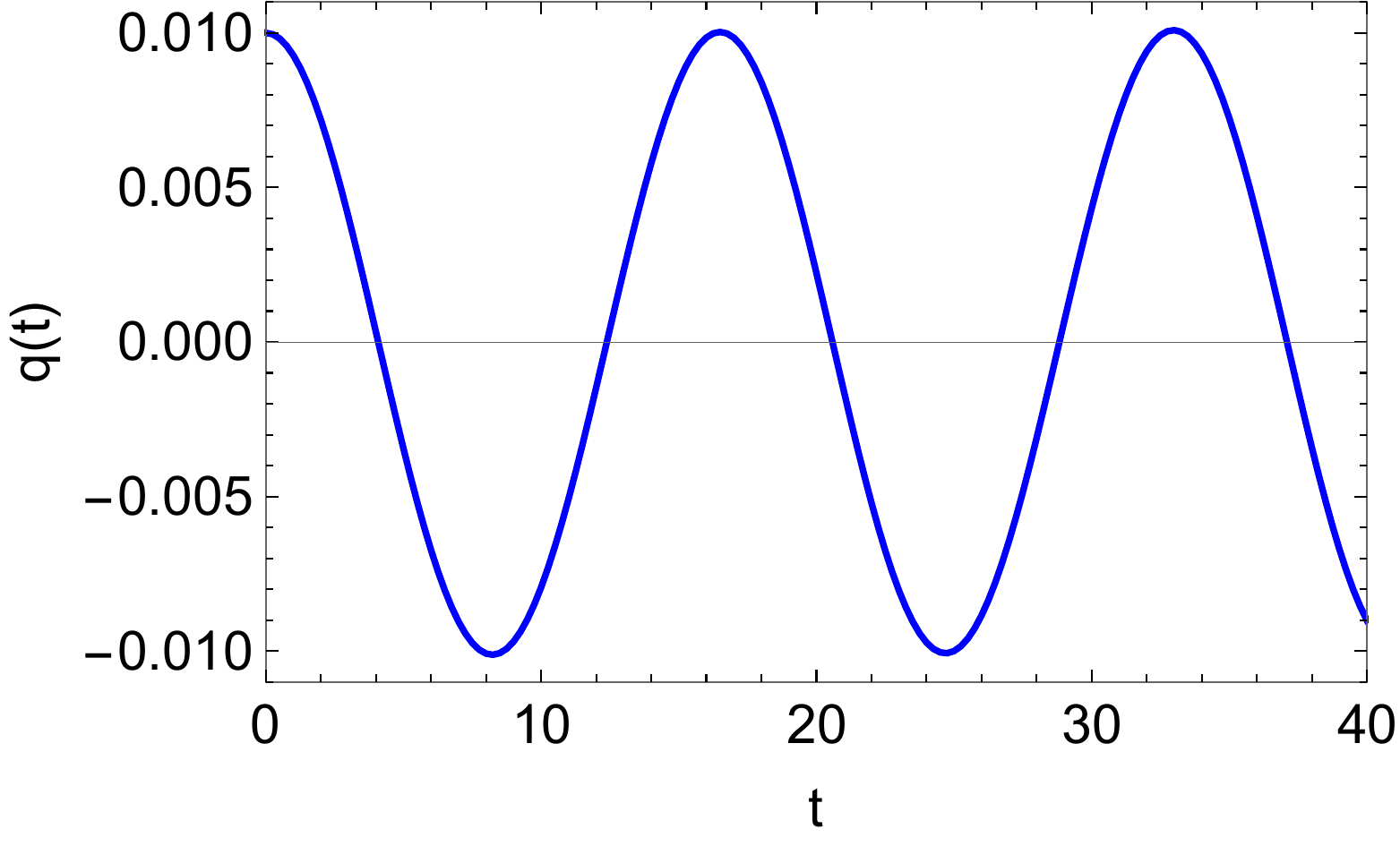} }
   \subfigure[\ $G(t)$ \vs\ $t$]
   { \label{f:Gt-case-IIIa}
     \includegraphics[width=0.9\columnwidth]{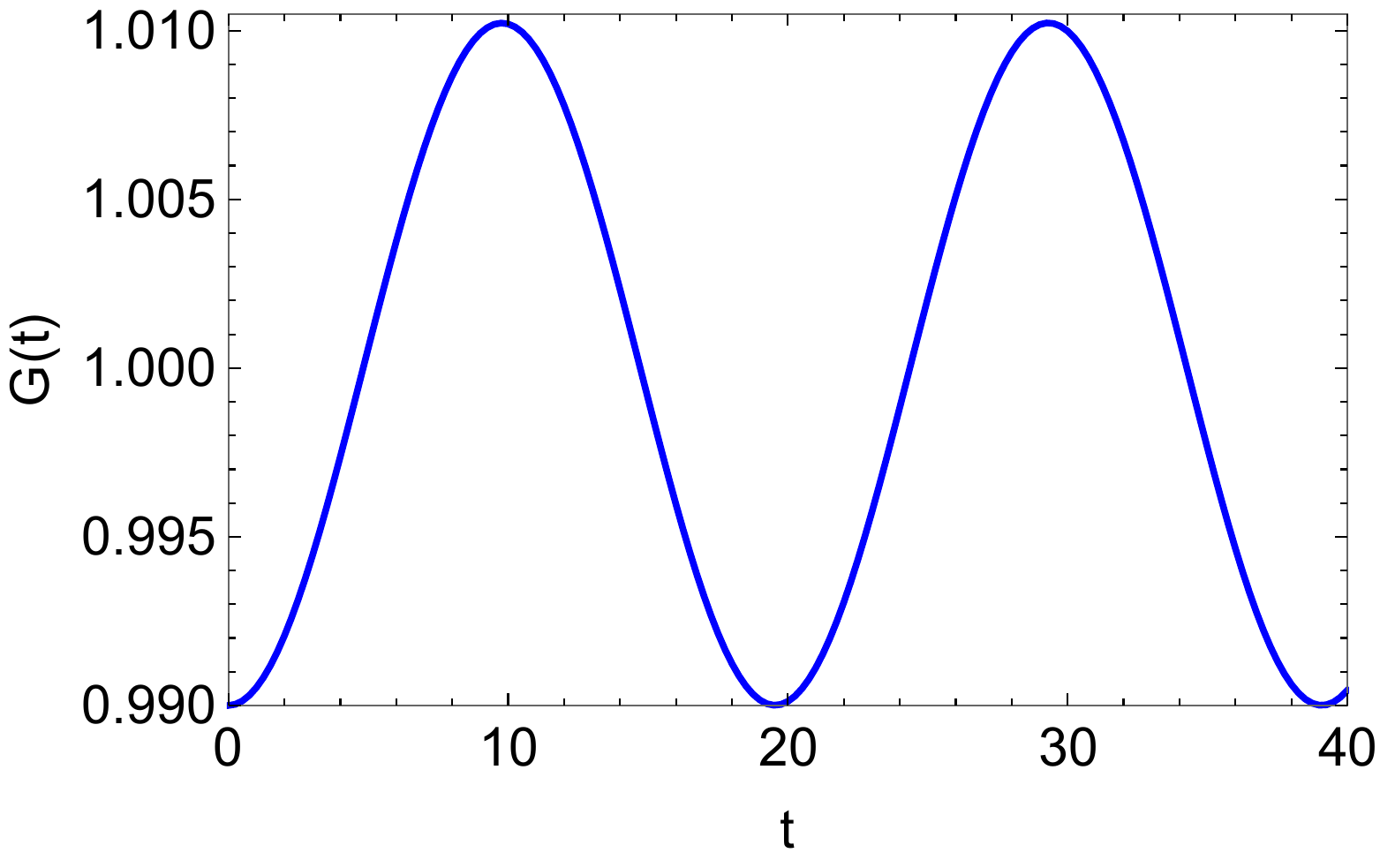} }
   \caption{\label{f:qGt-case-IIIa} Case III for $\kappa=3$ and $\tb^2 = 1/2$.
   Initial conditions are $G(0)=0.99$, $G'(0) = 0$,
   $q(0) = 0.01$, and $p(0) = 0$.}
\end{figure*}

For this case, all allowed solutions $\tb^2 > (\kappa+1)/\kappa^2$ should be stable to small changes in both $G$ and $q(t)$.  This oscillatory behavior for the case $\kappa=3$, $\tb^2=1/2$,  $q[0] = 0.01$ is  shown in Figs.~\ref{f:qGt-case-IIIa}.

%
%
%
%

%
%
\section{\label{s:numerics}Numerical Study of Stability}

%
%
\subsection{Domains of stability }

In order to study the stability of the soliton solutions, the actual
numerical simulation of the soliton evolution has been performed. For
that purpose, we have numerically solved Eq.~\ef{e:NLSE-1} with the
initial conditions described in Section~\ref{s:Model} using a Crank-Nicolson scheme \cite{ref:numrec}. 

The complex soliton in the spatial domain was represented on a regular grid with mesh size $\Delta x=5\times 10^{-3}$, and free boundary conditions were imposed.  

In order to study the stability regimes, we calculate the normalized correlation between the initial intensity profile (at $t=0$), and the intensity profile for $t>0$, \ie
\begin{equation}
   \text{corr}(0,t)
   =
   \frac{\int\limits_{-\infty}^{\infty} \dd{x}
   |\psi(x,0)|^2 \, |\psi(x,t)|^2}
   {\int\limits_{-\infty}^{\infty} \dd{x} |\psi(x,0)|^4} \>.
\end{equation}
Notice that $0\leq\text{corr}(0,t) = C_t < \infty$ and its value $C_t$ can be interpreted as follows
\begin{equation*}
   \text{corr}(0,t)
   =
   \begin{cases}
      C_t = 1 & \text{stable regime,}\\ 
      C_t>1   & \text{blow-up regime,}\\
      0\leq C_t<1 & \text{translational instability,}
   \end{cases}
\end{equation*}
for any $t>0$. Theoretically, it means that the evolution of the soliton solutions should be checked up to $t\to \infty$.  However, in practice, the inherent numerical noise of the simulations randomly perturbs the soliton shape during evolution, so finite evolution times are enough to determine the stability of soliton solutions. In this regard, we have found that the evolution of solitons up to $5\times 10^{2}$
time units with step size $\Delta t=3\times 10^{-2}$ is enough to study their stability.  
First let us return to the problem we studied earlier (Case I) where $\lambda=g=1$.  For that case three  methods predicted the same width stability region, namely
Derrick's theorem, setting $\omega_G^2 =0$ and the V-K criterion. In this case, if we consider the domain of numerical stability using the exact initial conditions, we agree with the results of Derrick's theorem.   

\begin{figure*}[t]  
   \centering
   \subfigure[\ Stability region (red dots) for an attractive potential 
   for case III with $\lambda = +1$ and $g=-1$.  In the empty region below 
   the solid line no soliton solution exists.]
   { \label{f:PS_attractive}
      \includegraphics[width=0.87\columnwidth]{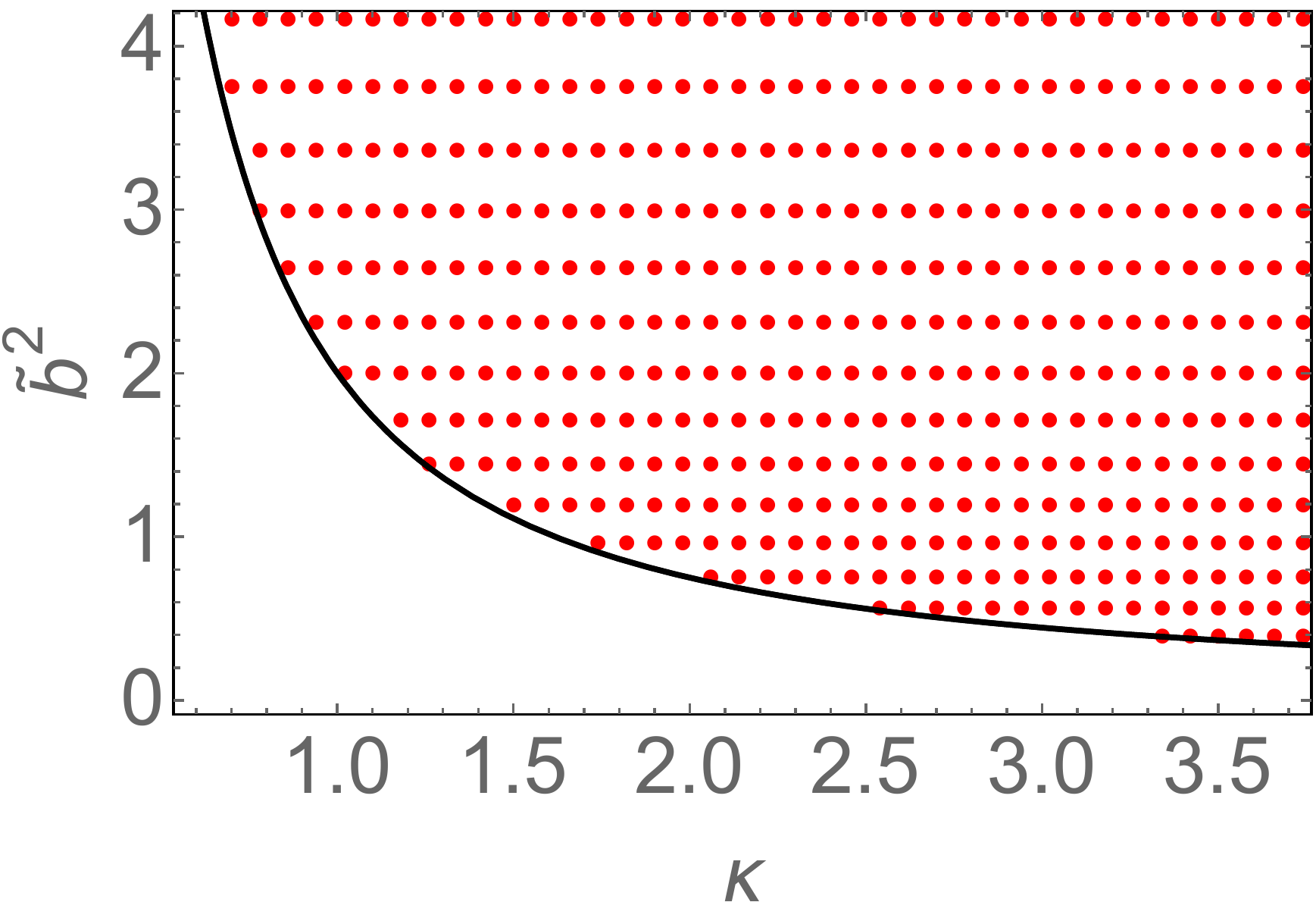} }
   \subfigure[\ Distribution of width instability regime (blue filled squares),
   and translational instability regime (yellow filled diamonds) for case II with 
   $\lambda=-1$ and $g=1$.]
   { \label{f:PS_repulsive}
      \includegraphics[width=0.95\columnwidth]{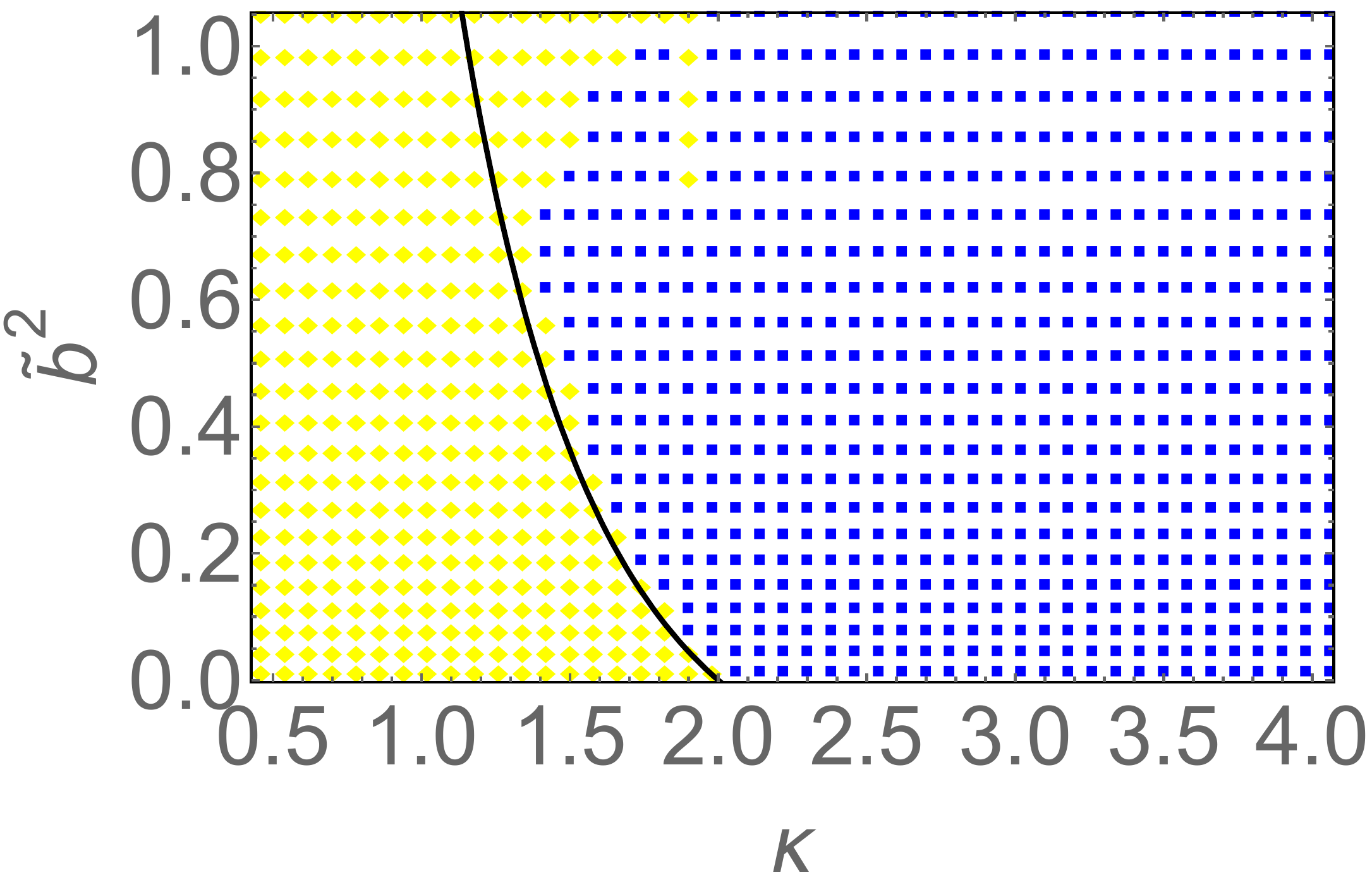} }
   \caption{\label{f:PS}Distribution of the stability 
   regions for (a) case III and (b) case II.}
\end{figure*}

Next look at the case when our variational method predicts stability, namely for the attractive potential with $g=-1$. The stability should occur as long as there is a solution namely $\tb^2 > (\kappa+1) /\kappa^2 $ as shown in Fig. (1a).  The numerical solutions lead to the same conclusions as shown in Fig. \ref{f:PS_attractive}.
Figures.~\ref{f:PS_attractive} and \ref{f:PS_repulsive} show the distribution of the stability region, blow-up regime, and translational instability for attractive and repulsive potentials, respectively.

%
%
%
%

In Fig. \ref{f:PS_repulsive} we show the regions of translational and width instabilities for Case II.

%
%
%
%

%
%
\subsection{\label{ss:trans}Effect of Translational Instability}

\begin{figure*}[t]
   \centering
   \subfigure[\ $\rho(x,t)$ at intervals of $\Delta t = 2$ for $q(0) = 0$.]
   { \label{f:trans-a}
     \includegraphics[width=0.95\columnwidth]{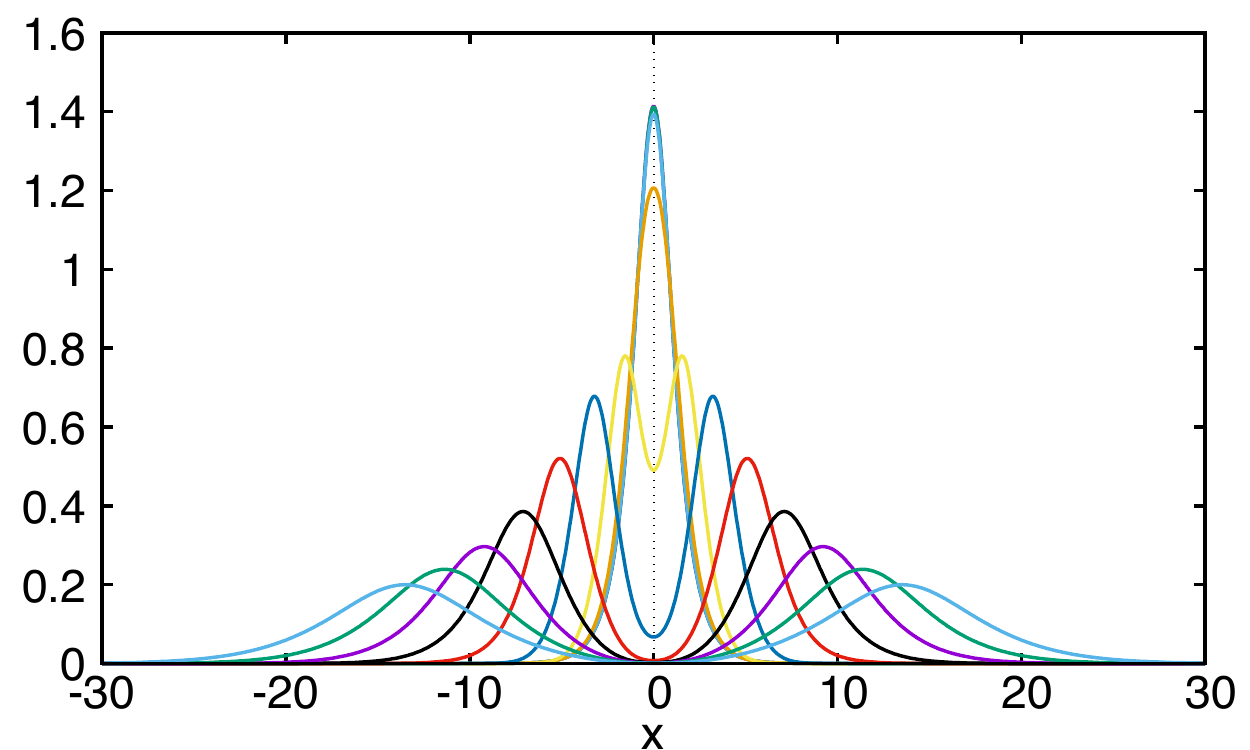} }
   \subfigure[\ $\rho(x,t)$ at intervals of $\Delta t = 2$ for $q(0) = 0.01$.]
   { \label{f:trans-b}
      \includegraphics[width=1.0\columnwidth]{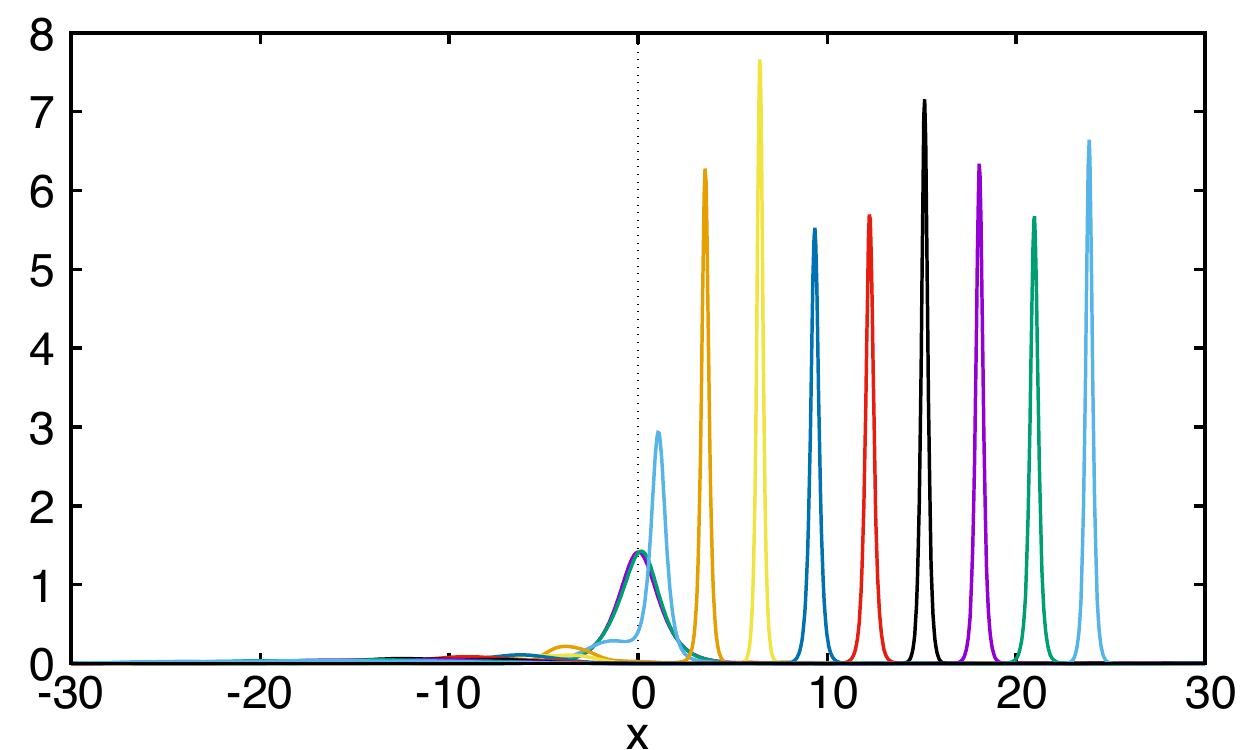} }
   \caption{\label{f:trans1}Density $\rho(x,t)$ calculated by a
   numerical solution of \Schrodinger's equation at intervals of $\Delta t = 2$
   for region II with $\kappa = 0.8$ and $\tb^2 = 1$, $G(0) = 1$,
   showing in (a) a bifurcation when $q(0) = 0$ of the initial wave function 
   into two waves going in opposite directions and in (b) initial condition 
   instability when $q(0) = 0.01$.}
\end{figure*}

This is illustrated for parameters in region II in Fig.~\ref{f:trans1} by numerical solutions of \Schrodinger's equation using a split-operator method.  In Fig.~\ref{f:trans-a}, the initial conditions are such that the solution ``rolls'' off the top of the potential in both directions and bifurcates, reminiscent of the quantum roll problem discussed in connection with the inflationary universe \cite{PhysRevD.34.3831}.  Although classically if we are at a maximum or saddle point we expect a particle to go to the right or to the left, since the underlying theory is essentially quantum mechanical, some of the mass density can go to the right or to the left.  If we start at the top of the saddle, and the repulsive potential is strong ($\tb^2 \sim 1$) then the solution bifurcates.  In Fig.~\ref{f:trans-b}, we give a slight kick away from the origin by setting $q(0) = 0.01$, which leads to translational motion away from the origin.  In this region of parameter space, the results are strongly dependent on the initial conditions.

%
%
%
%

%
%
%
%

%
%
\subsection{\label{ss:ICeffects}Effects of initial conditions}

\begin{figure}
   \centering
   \includegraphics[width=1.1\columnwidth]{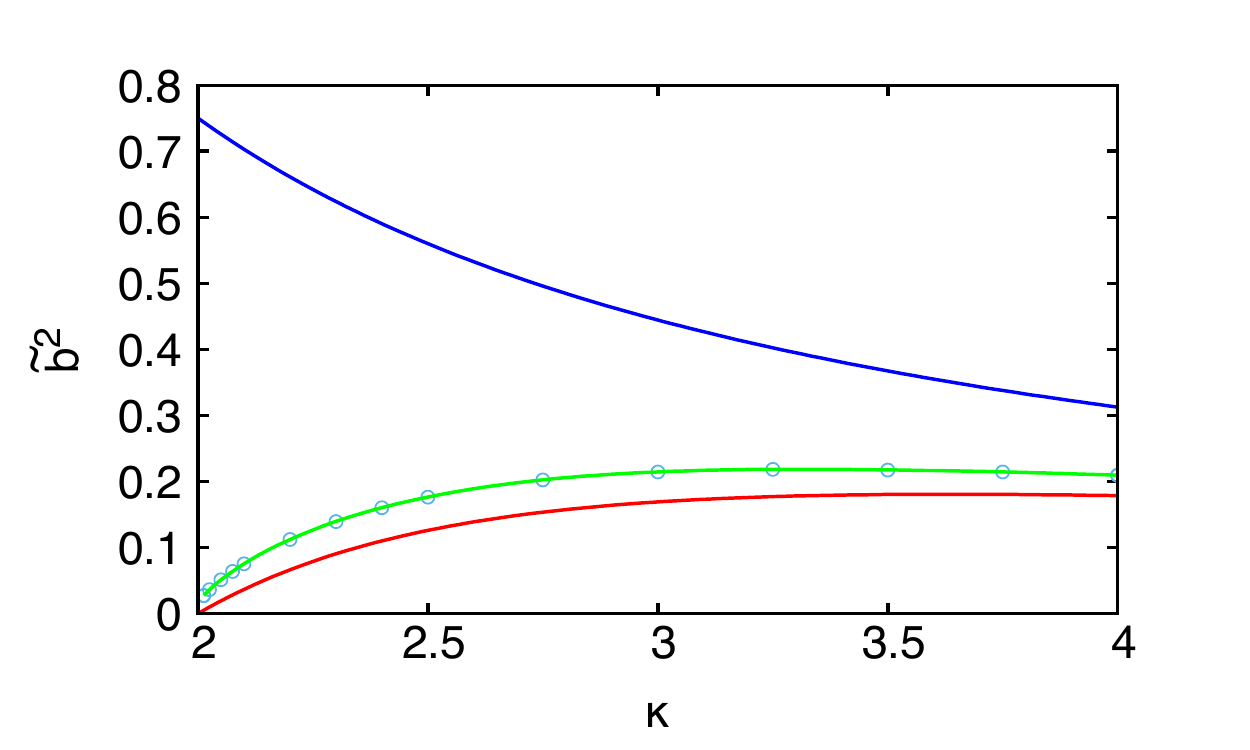}
   \caption{\label{f:IC-1} The circles are numerical calculations
   of ${\tilde b}^2_{\text{crit}}$ for $G(0)=1.01$, $q(0)=0$.  The
   green curve fit to these points lies \emph{above} the curve
   from Derrick's theorem (red line).  The upper curve (blue line) is
   $\tb_{\gamma}^2$.}
\end{figure}

A small perturbation of the initial conditions can also lead to instabilities.  These effects smear out the instability regions predicted by Derrick's theorem as shown in Fig.~\ref{f:Fig1}.  As an illustration of such effects, we show in Fig.~\ref{f:IC-1} the modification of region I for $\kappa > 2$ as a result of a 1\% change in $G(0)$ (using $G(0) = 1.01$)  instead of the exact solution value of $G(0) = 1.00$.  The unstable regime found by a numerical solution of \Schrodinger's equation with this initial condition is somewhat bigger than that found from Derrick's theorem (see Fig.~\ref{f:IC-1}).  Similar effects are found for other initial conditions.

%
%
%
%

\section{\label{s:conclude}Conclusions}

In this paper we studied the stability of two new exact classes of solutions of the NLSE in a real P{\"o}schl-Teller potential which is the SUSY partner of a \emph{complex} $\PT$-symmetric potential studied previously \cite{PhysRevE.92.042901}. Since the exact equations are derivable from an action principle,
if we approximate the wave function by a set of collective coordinates we obtain a symplectic formulation of the dynamics of the collective coordinates with
a conserved Hamiltonian. This reduced system is amenable  to several  approaches to studying the stability problem such as Derrick's theorem, looking at the energy landscape as a function of translations, as well as a  time dependent variational approach. Using these methods  we have mapped the domain of stability of these exact solutions and found good agreement with what we obtain from numerical simulations.   For the case of the solution with an attractive potential  but $g = -1$, the solutions are stable to both small width and small position deformations for all $\kappa$ as long as $\tb^2 > \gamma (\gamma+1)$. In that situation, the small oscillation equations of the variational approach give good agreement with numerical simulations. For the more interesting case of $g =1$ and a repulsive potential, there is a translational instability.   This can be seen by looking at the energy landscape or by looking at the small oscillation equations of our variational approximation.
The  stability results obtained from looking at the energy landscape of the solution as a function of both width stretching and translations are in agreement  with the results of a more rigorous linear stability analysis.   For the case of the repulsive potential which has a translational instability we find by performing numerical simulations quite interesting results.  If one starts with the exact initial solution,  the solution breaks into two equal amplitude pulses going in opposite directions. By perturbing the solution in one direction, the majority of the wave goes in that direction, but still some of the wave goes in the opposite direction.

%
%
\acknowledgments
F.C.~would like to thank the Santa Fe Institute  and the Center for Nonlinear Studies at Los Alamos National Laboratory for their hospitality. 
A.K.~is grateful to Indian National Science Academy (INSA) for awarding him INSA Senior Scientist position at Savitribai Phule Pune
University, Pune, India.
B.M.~and J.F.D.~would like to thank the Santa Fe Institute for their hospitality.  
B.M.~acknowledges support from the National Science Foundation through its employee IR/D program.
The work of A.S.~was supported by the U.S.~Department of Energy. 
E.A.~gratefully acknowledges support from the Fondo Nacional
de Desarrollo Científico y tecnológico (FONDECYT) project No. 1141223
and from the Programa Iniciativa Científica Milenio (ICM) Grant No. 130001.
The research of A.C.~was carried out at the Institute for Information Transmission Problems, Russian Academy of Sciences at the expense of the Russian Foundation for Sciences (Project 14-50-00150).
The work of R.L.~is supoported by the Department of Mathematics at Texas A\&M University.
%
%
\appendix
%
%
\section{\label{s:integrals}Useful integrals and definitions}

\begin{equation}\label{e:IT-1}
   c_1[\gamma]
   =
   \tint \dd{z} \sech^{2\gamma}(z)
   =
   \frac{\sqrt{\pi} \, \GammaF{\gamma}}{\GammaF{\gamma + 1/2}} \>.
\end{equation}
A useful result is
\begin{equation}\label{e:IT-2}
   c_1[\gamma+1]
     =
   \frac{2\gamma}{2\gamma+1} \, c_1[\gamma]
   \qc
   \frac{c_1[\gamma+1]}{c_1[\gamma]}
   =
   \frac{2\gamma}{2\gamma+1} \>.
\end{equation}
In Section~\ref{s:Derrick}, we defined an integral,
\begin{equation}\label{e:IT-3}
   g_1[\beta,\gamma]
   =
   \tint \dd{z} \sech^{2\gamma}(z) \, \sech^{2}(z/\beta) \>.   
\end{equation}
The first derivative of $g_1[\beta,\gamma]$ with respect to $\beta$ evaluated at $\beta=1$ is given by
\begin{align}\label{e:IT-4}
   \pdv{g_1[\beta,\gamma]}{\beta} \Big |_{\beta=1}
   &=
   2 \tint \dd{z} z \, \sech^{2\gamma+2}(z) \tanh(z)
   \\
   &=
   \frac{1}{\gamma+1} \tint \dd{z} \sech^{2\gamma+2}(z)
   =
   \frac{c_1[\gamma+1]}{\gamma+1} \>,
   \notag
\end{align}
where we have integrated by parts.  The second derivative, evaluated at $\beta = 1$ is
\begin{align}\label{e:IT-5}
   &\pdv[2]{g_1[\beta,\gamma]}{\beta} \Big |_{\beta=1}
   \\
   & \quad
   =
   2 \tint \dd{z}
   [\,
      - 
      z^2 \sech^{2\gamma+2}(z)
      -
      2 z \sech^{2\gamma+2}(z) \tanh(z)
      \notag \\
      & \qquad\qquad
      +
      3 z^2 \sech^{2\gamma+2}(z) \tanh^2(z) \,
   ] \>.
   \notag
\end{align}
The last term can be written as
\begin{equation*}
   \sech^{2\gamma+2}(z) \tanh^2(z)
   =
   \sech^{2\gamma+2}(z) - \sech^{2\gamma+4}(z) \>.
\end{equation*}
So \ef{e:IT-5} becomes
\begin{align}\label{e:IT-7}
   &\pdv[2]{g_1[\beta,\gamma]}{\beta} \Big |_{\beta=1}
   \\
   & \quad
   =
   2 \tint \dd{z}
   [\,
      2 z^2 \sech^{2\gamma+2}(z)
      -
      3 z^2 \sech^{2\gamma+4}(z)
      \notag \\
      & \qquad\qquad
      -
      2 z \sech^{2\gamma+2}(z) \tanh(z) \,
    ] \>.
   \notag   
\end{align}
Using \ef{e:DT-10.1}, Eq.~\ef{e:IT-7} becomes
\begin{equation}\label{e:IT-9}
   \pdv[2]{g_1[\beta,\gamma]}{\beta} \Big |_{\beta=1}
   =
   4 \, c_2[\gamma+1] 
   - 
   6 \, c_2[\gamma+2] 
   - 
   \frac{2 \, c_1[\gamma+1]}{\gamma+1} \>.
\end{equation}
We use this result in Section~\ref{s:Derrick}.
A second form of the critical curve can be found using an identity for $c_2[\gamma+1]$, which we derive here.  First note that
\begin{align}\label{e:IT-10}
   &c_2[\gamma+1]
   =
   \tint \dd{z} z^2 \, \sech^{2\gamma+2}(z)
   \\
   & \quad
   =
   \tint \dd{z} z^2 \, \sech^{2\gamma}(z) \,
   [\, 1 - \tanh^2(z) \, ]
   =
   c_2[\gamma]
   -
   I[\gamma] \>,
   \notag 
\end{align}
where
\begin{equation}\label{e:IT-11}
   I[\gamma]
   =
   \tint \dd{z} z^2 \, \sinh^2(z) \, \sech^{2\gamma+2}(z) \>.
\end{equation}
Using  the identity,
\begin{align}
   \pdv[2]{\sech^{2\gamma}(\lambda z)}{\lambda}
   &=
   2 \, \gamma (2 \gamma + 1) \, z^2 \, \sinh^2(\lambda z) \, 
   \sech^{2\gamma+2}(\lambda z)
   \notag \\
   & \qquad
   -
   2 \, \gamma \, z^2 \, \sech^{2\gamma}(z) \>,
   \label{e:IT-12}
\end{align}
integrating it over $z$, and evaluating at $\lambda \rightarrow 1$ gives:
\begin{equation}\label{e:IT-13}
   2 \, c_1[\gamma]
   =
   2 \, \gamma (2 \gamma + 1) \, I[\gamma]
   - 
   2 \, \gamma \, c_2[\gamma] \>.
\end{equation}
Substituting in $I[\gamma]$ from Eq.~\ef{e:IT-10} gives:
\begin{equation}\label{e:IT-14}
   c_2[\gamma+1]
   =
   \Bigl ( \frac{2 \gamma}{2 \gamma + 1} \Bigr ) \, c_2[\gamma]
   -
   \Bigl ( \frac{1}{\gamma ( 2 \gamma + 1)} \Bigr ) \, c_1[\gamma] \>.
\end{equation}
Using this result, and after a bit of algebra, it is easy to show that
the critical curve Eq.~\ef{e:DT-12} can be written as
\begin{equation}\label{e:IT-15}
   \tb^2_{\text{crit}}
   =
   - \lambda \,
   \frac{\gamma(2\gamma-1)}
   {\displaystyle
      \frac{(\gamma - 1)(2\gamma - 1)}{(\gamma + 1)(2\gamma + 3)}
      +
      \frac{4 \gamma^2}{2\gamma + 3} \, \frac{c_2[\gamma]}{c_1[\gamma]} } \>.
\end{equation}
This form is identical to Eq.~(4.28) in Ref.~[\onlinecite{Cooper:2017aa}].

%
%
%
\newpage
\bibliography{johns.bib}
%
%
\end{document}